# Production of Ammonia Makes Venusian Clouds Habitable and Explains Observed Cloud-Level Chemical Anomalies.

William Bains[1,2], Janusz J. Petkowski[1], Paul B. Rimmer[3,4,5], Sara Seager[1,6,7,*]

[1]Department of Earth, Atmospheric, and Planetary Sciences, Massachusetts Institute of Technology, Cambridge, MA 02139, USA
[2]School of Physics & Astronomy, Cardiff University, 4 The Parade, Cardiff CF24 3AA, UK
[3]Department of Earth Sciences, University of Cambridge, Downing St, Cambridge CB2 3EQ, United Kingdom
[4]Cavendish Laboratory, University of Cambridge, JJ Thomson Ave, Cambridge CB3 0HE, United Kingdom
[5]MRC Laboratory of Molecular Biology, Francis Crick Ave, Cambridge CB2 0QH, United Kingdom
[6]Department of Physics, Massachusetts Institute of Technology, Cambridge, MA 02139, USA
[7]Department of Aeronautics and Astronautics, Massachusetts Institute of Technology, Cambridge, MA 02139, USA.
\*  Corresponding author Sara Seager
**Email:**  seager@mit.edu

## Abstract

The atmosphere of Venus remains mysterious, with many outstanding chemical connundra. These include: the unexpected presence of ~10 ppm $O_2$ in the cloud layers; an unknown composition of large particles in the lower cloud layers; and hard to explain measured vertical abundance profiles of $SO_2$ and $H_2O$. We propose a new hypothesis for the chemistry in the clouds that largely addresses all of the above anomalies. We include ammonia ($NH_3$), a key component that has been tentatively detected both by the Venera 8 and Pioneer Venus probes. $NH_3$ dissolves in some of the sulfuric acid cloud droplets, effectively neutralizing the acid and trapping dissolved $SO_2$ as ammonium sulfite salts. This trapping of $SO_2$ in the clouds together with the release of $SO_2$ below the clouds as the droplets settle out to higher temperatures, explains the vertical $SO_2$ abundance anomaly. A consequence of the presence of $NH_3$ is that some Venus cloud droplets must be semi-solid ammonium salt slurries, with a pH~1, which matches Earth acidophile environments, rather than concentrated sulfuric acid. The source of $NH_3$ is unknown, but could involve biological production; if so, then the most energy-efficient $NH_3$-producing reaction also creates $O_2$, explaining the detection of $O_2$ in the cloud layers. Our model therefore predicts that the clouds are more habitable than previously thought, and may be inhabited. Unlike prior atmospheric models, ours does not require forced chemical constraints to match the data. Our hypothesis, guided by existing observations, can be tested by new Venus in situ measurements.

## Significance Statement

This research provides a transformative hypothesis for the chemistry of the atmospheric cloud layers of Venus while reconciling decades-long atmosphere anomalies. Our model predicts that the clouds are not entirely made of sulfuric acid, but partially composed of ammonium salt slurries, which may be the result of biological production of ammonia in cloud droplets. As a result, the clouds are no more acidic than some extreme terrestrial environments that harbor life. Life could be making its own environment on Venus. The model's predictions for the abundance of gases in Venus' atmosphere matches observation better than any previous model, and are readily testable.

**Main Text**

**Introduction**

Venus is often called Earth's sister planet because of its similar mass and size to Earth. Yet owing in part to the greenhouse effect from its massive $CO_2$ atmosphere, Venus's surface temperature is higher than 700 K—too hot for life of any kind. The Venusian surface is therefore a complete contrast to Earth's temperate surface and rich surface biosphere. Nonetheless, scientists have been speculating on Venus as a habitable world for over half a century (1–7). Such speculations are based on the Earth-like temperature and pressure at the altitudes of 48-60 km above the surface (8, 9).

Venus' is perpetually shrouded in a ~20 km-deep layer of clouds, including the temperate atmosphere layers at 48-60 km. The prevailing consensus is that the clouds of Venus are made from droplets of concentrated sulfuric acid. This conclusion is inferred from the presence of small amounts of sulfuric acid vapor in the atmosphere (10, 11) and the refractive index of cloud droplets (12, 13). While the clouds are often described as 'temperate' or 'clement', such a statement is misleading when it comes to habitability. If the cloud particles are actually made of concentrated sulfuric acid then it is difficult to imagine how life chemically similar to life on Earth could survive (7, 14). Specifically, the aggressive chemical properties of sulfuric acid and the extremely low atmospheric water content (14, 15) are orders of magnitude more acidic and 50-100 times drier than any inhabited extreme environment on Earth.

**Overview of Venusian Atmosphere Anomalies**

Despite over 50 years' of remote and local observation, Venus' atmosphere has a number of lingering anomalies with either poor model fits or no explanations (16, 17).

One such long-standing mysterious feature of the atmosphere, which is not well explained by current atmospheric chemistry models, is the abundance profile of water vapor and $SO_2$ in and above the cloud layers. (17–19).

Observations show that $H_2O$ persists throughout the atmosphere while the $SO_2$ is observed in ppm abundances below the clouds and ppb abundance above the clouds. Yet, expectations are very different. The primary source of $SO_2$ and $H_2O$ in the atmosphere of Venus is volcanism. As the gases are released from volcanoes they are uniformly mixed vertically throughout the atmosphere. At very high altitudes in the atmosphere, around 70 km, $SO_2$ and $H_2O$ are efficiently destroyed by UV radiation. However the observed $SO_2$ and $H_2O$ abundance profiles deviate from the uniform distribution, notably that $SO_2$ shows significant depletion in the cloud layers and $H_2O$ is present above the cloud layers.

Previous consensus models explained the $SO_2$ profile by suggesting that $SO_2$ is photochemically oxidized to $SO_3$, which then reacts with water to form sulfuric acid in the clouds:
$CO_2 + h\nu \rightarrow CO + O$
$SO_2 + O + M \rightarrow SO_3 + M$
$SO_3 + 2H_2O \rightarrow H_2SO_4 + H_2O$
$SO_2 + H_2O + CO_2 \rightarrow H_2SO_4 + CO$

However, as there is 5 x more $SO_2$ than $H_2O$, this chemistry should strip all the water out of the cloud layer, and additionally react with and prevent water from reaching and accumulating above the clouds as well, while only reducing $SO_2$ by 20%, not the 99.9% observed (20). Previous models provide a numerical fix to match the observations, arbitrarily removing SO$_2$ or artificially keeping the water abundance constant (21, 22).

Another mystery is the presence of O$_2$ in the clouds (23, 24), as there is no known process for O$_2$ formation in the cloud layers (discussed further below). Finally, the SO$_2$, O$_2$, and H$_2$O anomalies, together with other trace atmospheric gas abundances, form a chemical disequilibrium in the clouds of Venus (25–27).

A more tentative but intriguing anomaly is that of the detection of NH$_3$ in and below the cloud layers. NH$_3$ was tentatively detected both by the Venera 8 chemical probe (28) and in reanalyzed Pioneer Venus (Pioneer 13) data (27). The re-analysis of Pioneer Venus data showed additional N species (NO$_x$), suggesting further chemical disequilibrium in the cloud layers. The cloud particles themselves also contain many unknowns. The largest particles, predominant in the lower cloud decks (called Mode 3 particles (29)), may have a substantial solid component, implying that they cannot be exclusively made of liquid concentrated sulfuric acid (30).

Some additional anomalies that are not directly relevant to this work, such as the 'unknown UV absorber' (31) and the possible presence of methane (32) or phosphine (33, 34), have all been suggested as signs of life in the clouds.

How the Rimmer et al. model resolves the SO$_2$ and H$_2$O abundance conundrum

Recently, Rimmer et al. 2021 (20) proposed a mechanism to explain the depletion of SO$_2$ in the atmospheric cloud layers, as well as the vertical abundance profile of H$_2$O in and above the clouds. If a base is present inside the cloud sulfuric acid droplets, SO$_2$ will dissolve in the liquid droplets (by reaction with OH$^-$) to form sulfite. The base (B), therefore, traps the SO$_2$ inside the cloud droplet as sulfite (HSO$_3^-$),

SO$_2$ + H$_2$O + B → SO$_2$ + BH$^+$ + OH$^-$ → BH$^+$ + HSO$_3^-$. 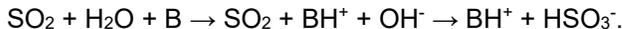

In summary, the equilibrium of the reaction

SO$_2$ + H$_2$O ↔ H$_2$SO$_3$ 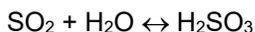

is pulled to the right of the above equation, and S(IV) species are trapped as sulfite salts through reaction with the base. Thus, SO$_2$ is depleted in the cloud layer, compared to the model with no bases. Eventually, the cloud droplets rain down to lower atmosphere layers and the salts dissociate due to higher temperatures, releasing SO$_2$.

Water is consumed in the sulfite-forming reaction, but is recycled into the lower atmosphere on breakdown of the sulfites, which provides a mechanism to explain the water vapor abundance profile through the clouds. Some water is removed from the cloud layer, but because it is replenished by recycling from below the clouds, the water removal is not absolute, and so some water remains at the cloud top and in the atmosphere above the clouds. Thus the Rimmer et al model predicts that both SO$_2$ and H$_2$O will be present above the clouds but at substantially lower abundance than they are below the clouds, in agreement with observation.

The formation of the sulfite salt within a droplet effectively neutralizes the acid in the droplet, with the very important outcome that some of the cloud droplets are much less acidic than previously thought, with a pH between -1 and 1 (20), instead of an acidity of approximately -11 (on the Hammett acidity scale). If correct, the revised pH range of some droplets has a significance for the habitability of the clouds of Venus that cannot be overstated. Such a pH range is habitable by terrestrial extremophiles (35), as compared to the acidity of concentrated sulfuric acid in which all terrestrial life, and most terrestrial biochemicals, would be destroyed (14).

We argue that the identity of any droplet-neutralizing base is unknown. Rimmer et al (20) adopted NaOH as a model base for their calculations, but noted that iron oxides were a more physically realistic possibility. In principle minerals that can absorb $SO_2$ could be delivered to the clouds from Venusian volcanic eruption, wind-lofting of dust, or from meteoritic infall. However it has not been demonstrated that such mechanisms could deliver the very high amount of ~20 tonnes per second flux of mineral salts (specifically iron oxides) required (20).

We are motivated to extend (20)'s analysis with the hypothesis that the neutralizing base that is capturing $SO_2$ is locally generated in the clouds. We postulate that $NH_3$ is the neutralizing agent for the Venusian cloud droplets, trapping $SO_2$ and thus explaining the drop in $SO_2$ abundance across the clouds. We are additionally motivated by the tentative in situ observations of $NH_3$ in the Venus cloud layers, both from Venera 8 chemical assay (28) and Pioneer Venus probe mass spectrometry (27). If present, $NH_3$ observations cannot yet be readily explained through any known abiotic planetary processes (36). We therefore also explore the possibility that the $NH_3$ is biologically-produced.

**Results**

Ammonia as a Neutralizing Agent in the Venusian Cloud Droplets

We propose $NH_3$ as the only plausible neutralizing base that can be generated in situ in the clouds from gas phase components (See SI Section 1 for further details on potential neutralizing agents in the cloud layers). The presence of $NH_3$, as with any neutralizing base, leads to chemistry that results in the $SO_2$ depletion in the clouds and the observed $H_2O$ abundance profile, and is consistent with a subset of Mode 3 particles being non-spherical (i.e., not liquid) and not composed of pure concentrated sulfuric acid. The presence of $NH_3$ may also solve the otherwise unexplained presence of $O_2$ in the clouds, especially if the source of $NH_3$ is biological.

To support our hypothesis that $NH_3$ could explain the presence of $O_2$ within the clouds, we first explore the limited number of possible chemical reactions that could lead to the formation of $NH_3$ in the Venusian atmosphere cloud-layer conditions (Table 1).

The most abundant source of nitrogen atoms in the atmosphere of Venus is $N_2$ gas, so to make $NH_3$, $N_2$ must be reduced to $NH_3$. The reduction of $N_2$ to form $NH_3$ requires a source of hydrogen atoms, and a source of electrons (reducing equivalents). Hydrogen atoms are rare in the atmosphere of Venus. The most abundant gas-phase source of hydrogen atoms in the atmosphere of Venus is $H_2O$, followed by HCl. In order to generate reducing equivalents, some species must be oxidized. Species available to be oxidized include CO, OCS, $SO_2$, $N_2$, $H_2O$, and HCl. Phosphorus, if present, will be overwhelmingly present as $H_3PO_4$ (34); neither $H_3PO_4$ or $CO_2$ can be further oxidized.

The most energy- and water-efficient $NH_3$-producing reaction (Reaction 3 in Table 1) also produces molecular oxygen. We choose Reaction 3: $2N_{2(aq)} + 10H_2O_{(l)} \rightarrow 4NH_4^+OH^-_{(aq)} + 3O_{2(aq)}$ as the basis for our model for two reasons, Firstly, parsimony leads us to prefer a reaction that uses the smallest amount of rare materials ($H_2O$ and energy). Secondly, Reaction 3 is the only $NH_3$-forming reaction that directly produces $O_2$ in the clouds (Table 1), whose detection is one of the anomalies we wish to explain (discussed below); the other reactions produce different oxidized species which would not be observed but which would also produce $O_2$ on breakdown, and at the cost of greater energy consumption.

**A key question is** what $NH_3$ production rate (by Reaction 3) is needed for maintaining the low $SO_2$ abundance, as compared to expected equilibrium values in the atmospheric cloud layers. We base the $SO_2$ production rate on the rate at which $SO_2$ would be replenished into the clouds by mixing from below, and hence the rate at which it must be removed from the clouds. The flux is ~$10^{11}$ tonnes per year $NH_3$, which is on the order of photosynthetic production of $O_2$ on Earth (see Methods). This flux is calculated assuming that $NH_3$ is only produced to sequester $SO_2$, and that only $NH_3$ sequesters $SO_2$. If other species contribute to removing $SO_2$, whether hydroxide salts, iron oxides, or other species, the $NH_3$ production will be accordingly lower. Any byproduct of $SO_2$ sequestration must have a flux of ~$10^{11}$ tonnes per year at the bottom of the clouds, based on the $SO_2$ depletion within the clouds. A flux of ~$10^{11}$ tonnes per year is consistent, to within an order of magnitude, to the mass loss at the bottom of the clouds from rainout of Mode 3 particles from our calculations (see SI Section 2).

All of the $NH_3$-producing reactions in the Venusian atmosphere conditions are highly endergonic (Table 1), and so must be coupled to an energy source if the reactions are to produce net, 'surplus' $NH_3$. There are several energy sources that could in principle drive the production of $NH_3$. Lightning falls short by many orders of magnitude of the necessary rate of production of $NH_3$ (See SI Section 7.1; Table S3), and is very unlikely to produce both $NH_3$ and $O_2$ simultaneously. Similarly, UV photochemistry is unlikely to produce $NH_3$ in more than trace amounts (see SI Section 7.2), although we note that the photochemistry of nitrogen species in concentrated sulfuric acid have not been explored. Volcanic sources of $NH_3$ on Earth are closely associated with organic deposits, including coal, and also are quantitatively insufficient, even based on terrestrial rates of volcanic $NH_3$ production, which are likely to be much higher than any plausible $NH_3$ production on Venus (see SI Section 7.3; Figure S4).

The ability to couple chemical energy to drive endergonic reactions is a universal characteristic of life, and specifically the use of energy to drive the reduction of $N_2$ to $NH_3$ in an oxidizing environment is widely found in terrestrial organisms (37, 38). We should therefore consider the possibility that living organisms in the clouds of Venus are making $NH_3$. All of the $NH_3$-producing reactions presented in Table 1 consume water, which is a rare resource in the clouds of Venus. The energy expended and water molecules consumed in the process of making $NH_3$ must be balanced by an equally powerful benefit to the organism for this apparently wasteful chemistry. Neutralizing the acid to make the droplets habitable is a clear benefit.

We discuss the other, possibly insuperable barriers to the concept of life in the Venusian clouds below. Here we only note that the presence of life could explain the observed presence of $NH_3$ and $O_2$, and later show that it could explain the observed vertical abundances of $H_2O$ and $SO_2$ within and above the atmospheric cloud layers, and the semi-solid nature of Mode 3 particles. An additional consequence of the $NH_3$ cloud-droplet chemistry is that the pH of cloud particles with dissolved $NH_3$ must have a pH between -1 and 1, as first shown by Rimmer et al. (20) for NaOH (Figure 2).

<u>The flux of $NH_3$ is within the plausible biomass production</u>

The flux of $NH_3$ needed to achieve the neutralization effect is not prohibitive for a realistic biomass within the cloud droplets. We calculate the biomass required by this model as follows. The production of $10^{11}$ tonnes per year is equivalent to $3 \cdot 10^9$ grams $NH_3$ per second. Several species of cyanobacteria fix nitrogen at an average rate of ~$4 \cdot 10^{-7}$ grams/gram wet weight biomass per second (39–41). If life is present in the clouds of Venus, it will not be terrestrial life; however if we take these terrestrial organisms as precedent, $10^{11}$ tonnes per year would be produced by ~$8 \cdot 10^{15}$ grams wet weight of organism. While this mass might appear significantly high, it is ~1/2000 (0.05%) the biomass of the Earth (42). This mass translates to ~1.5% of the mass of cloud particles in the lower 5 km of the cloud deck (25).

Our model for the production of $NH_3$ by life is summarized in Figure 1.

Towards a resolution of Venus atmospheric anomalies

The incorporation of $NH_3$ in our photochemistry model of the Venusian atmosphere produces profiles of atmospheric gases that match the observed abundances of some atmospheric gases better than existing models of Venus's atmosphere. Although $NH_3$ is an input to our model, no existing Venus photochemical models include $NH_3$ (e.g., (22, 43)). In Figures 3 and 4 we show a summary of the output of the modeling with $NH_3$ included, compared to the same model run without $NH_3$ and $O_2$ input, the latter as reported in (20). The atmospheric photochemistry of the clouds was modeled as described in (20, 34), and is summarized in Methods. Specifically, our model better explains compared to previous models: 1) the observed disequilibria in the clouds of Venus; 2) the measured, but subsequently ignored, abundances of $O_2$ in the clouds; 3) the abundance profile of water vapor; 4) the tentative detections of $NH_3$ by Venera 8 and Pioneer Venus probes; and the abundance profile of $SO_2$ through the cloud layers. To demonstrate how well our new model with $NH_3$ fits the measured data, we show three model results in Figures 3 and 4: one model with $NH_3$; one model without $NH_3$ but with an unphysical arbitrary depletion rate of $SO_2$ (a fix common amongst other models in order to fit the data); and one model without $NH_3$ and without any artificial chemical constraints.

We now turn to each relevant atmosphere anomaly, first reviewing the data, then how the presence of $NH_3$ helps resolve the anomaly.

*$O_2$ in the clouds is a natural outcome of $NH_3$ production*

Our model provides an explanation for the presence of $O_2$ in the Venus cloud layers. $O_2$ has been measured via *in situ* measurements (44, 45). The Pioneer Venus Gas Chromatography (GC) reported 43.6 ppm molecular oxygen ($O_2$) in the clouds at 51.6 km, 16 ppm below the clouds at 41.7 km, and no detection of oxygen at 21.6 km (23). The Venera 14 GC detected 18 ppm $O_2$ average between 35 and 58 km (24). (The Neutral Mass Spectrometer (LNMS) on Pioneer Venus showed a signal of 32 amu, but this was attributed to $O_2$ ions formed from reaction of $CO_2$ in the mass spectrometer (46), therefore was considered unreliable. However, we emphasize that this uncertainty about the source of $O_2$ is specific to mass spectrometry (47)). We also note that several ground-based observations attempted to provide upper limits for the abundance of $O_2$ above the clouds (48, 49). The spectroscopic searches for $O_2$ have been subjected to varying interpretations (16, 17) and are claimed to be difficult to reconcile with the in-cloud $O_2$ abundance detected by both Pioneer Venus and Venera probes because one expects to observe a gradient of $O_2$ from above to below the clouds. Such discrepancies can only ultimately be resolved by new in situ measurements of $O_2$ in the clouds of Venus.

In the past, the validity of $O_2$ has been challenged based on thermodynamics. Initial studies of the atmosphere of Venus in the 1970s and 1980s assumed the atmosphere was at thermodynamic equilibrium. One author discounted $O_2$ as follows (44), "*We therefore conclude, that either we have to accept a strong disequilibrium state among CO, $SO_2$, $O_2$ and $H_2O$ in the lower atmosphere of Venus, or discard at least one of the measurements in order to save the assumption of thermodynamic equilibrium. The latter course is our preferred one.*" Some subsequent studies followed this argument (17, 23, 36), although not all (50), and the author himself modified his opinion in a subsequent paper (45). By now it has been accepted for over two decades that the atmosphere of Venus is not at thermodynamic equilibrium (25, 26, 51, 52), although Venus' atmosphere is not as far from disequilibrium as Earth's atmosphere is (51, 52). Recently, the re-analysis of the Pioneer Venus data showed the atmosphere was further from equilibrium than previously thought, due the presence of a range of reduced gases (27). Still, the cause of the Venus atmosphere thermodynamic disequilibrium is one of the unsolved problems in Venus science (17).

If the chemistry of $NH_3$ production is the source of $O_2$, then our model predicts on order 1 ppm $O_2$ in the cloud level between about 50 to 60 km. 1 ppm is 20-fold lower than the measured values (23, 24). However, the value of 1 ppm at lower altitudes is far greater (15 orders of magnitude) than predicted by our and other photochemistry models that exclude $NH_3$. While there are no known non-biological processes that could produce $O_2$ locally in the clouds of Venus, we note for future work that other biological processes such as oxygenic photosynthesis could also be contributing to the overall $O_2$ budget in the clouds.

It has been suggested that $O_2$ could also be produced by lightning, which is consistent with $O_2$'s presence in and below but not above the clouds (53). Lightning and coronal discharge can produce $O_2$ in a $CO_2$ + $N_2$ atmosphere (54). A thermodynamic-based calculation suggests that the amount of $O_2$ possibly produced by lightning is 4-5 orders of magnitude too low to explain the observations (SI Section 8.1; Table S4). However, the efficiency of the production of $O_2$ by lightning could be tested experimentally on Earth. It is possible that all the $O_2$ detections summarized above were made as spacecraft fell through high-intensity storm regions (55), but it seems an unlikely coincidence for two or three separate probes to experience storms. In addition, any $NH_3$ present in the clouds would be destroyed by the lightning and only trace amounts would reform (SI Section 7.1). The thermal decomposition of $H_2SO_4$ to $O_2$ and $SO_2$ has been suggested as an industrial process (56) but it is unlikely under Venus conditions (see SI Section 8.2; Figure S5).

At altitudes above the cloud level (~62 km), no $O_2$ has been detected, strongly suggesting a fractional abundance of less than $10^{-7}$ (48). Yet, all existing photochemical models predict significant molecular oxygen above the clouds (e.g., (22)) due to the instability of $CO_2$ to photolysis. $CO_2$ is dissociated into CO and O, which cannot rapidly recombine because the recombination reaction is spin forbidden. Some alternative pathway, involving, for example, OH chemistry, sulfur chemistry or chlorine chemistry is required to restore $CO_2$ (see (18)), but none of these pathways are sufficient to draw above-cloud $O_2$ below 1 ppm (22). This mismatch between the extremely low observed $O_2$ levels above the clouds and the higher predicted levels is a well-known conundrum of Venus' cloud layer chemistry. Our model provides a partial solution by predicting a reduced $O_2$ level above the clouds compared to the same model without $NH_3$ (Figures 3 and 4).

### Model output $H_2O$ and $SO_2$ abundance profiles are consistent with observations

Our photochemistry model with $NH_3$ production is, together with the model it is based on (20), consistent with the observed $H_2O$ and $SO_2$ abundance profiles in and above the clouds.

$SO_2$ and $H_2O$ have been observed on many occasions by remote campaigns, orbiters and in situ probes (reviewed in (20, 26)). For example, the Visible and Infrared Thermal Imaging Spectrometer (VIRTIS) instrument on board Venus Express observed a mean abundance of $H_2O$ and $SO_2$, below the clouds at 30-40 km, to be ~30 ppm and ~150 ppm respectively (57). The observed abundances of $H_2O$ and $SO_2$ just below the clouds are consistent between remote, orbiter, and in situ observations (20). Recall, that the 5 x excess $SO_2$ over $H_2O$ should strip all the water out of the cloud layer, and hence remove all water above the clouds as well, a solution that is not consistent with observations. The Rimmer et al. model (20) uses cloud chemistry ($NH_3$ or mineral bases) to strip the $SO_2$ in the clouds. As a result water remains in the clouds and above the clouds which agrees with the remote, orbiter and in situ observations of a few ppm of $H_2O$ above the cloud layers (reviewed in (20)).

Within the Venus cloud layers, there is substantial difference amongst measurements of water abundance in the clouds as summarized by (20), which may represent varied local conditions.

In previous models to the one described here, water vapor is removed at the cloud tops by reaction with $SO_3$ to form sulfuric acid, which then condenses out to form the cloud droplets. Since there is more below-cloud $SO_2$ than $H_2O$, all the $H_2O$ above the clouds is be removed, some models even showing complete depletion of $H_2O$ (33). Yet, this depleted $H_2O$ above the clouds does not match observations which show plenty of water vapor above the clouds (Figures 3 and 4). Previous models to ours solve this problem with physically unrealistic numerical fixes; either including excess $H_2O$ below the clouds (21), or fixing the $H_2O$ abundance to observed values, such that any reactions involving $H_2O$ do not consume any $H_2O$ (22). Most models avoid the water vapor abundance problem altogether by restricting the calculations only to a section of the atmosphere, above the clouds or below the clouds.

Critically important is that our model without $NH_3$ (33) must, similarly to other models, impose non-physical constraints on $SO_2$ chemistry in order to make the $SO_2$ gas abundance profile fit observations, specifically by adding an arbitrary removal rate for $SO_2$ in the clouds tuned to fit the data.

We emphasize that our model that includes $NH_3$ or another base (20) is the only model known that avoids artificial fixes of $SO_2$ and $H_2O$. To further emphasize this point, Figures 3 and 4 include our very poorly fitting model gas abundance profiles without $NH_3$ and without the artificial $SO_2$ removal rate.

Below the clouds, our photochemical model with $NH_3$ predicts the same $H_2O$ abundance as models without $NH_3$, including previous models (e.g., (43)).

*$NH_3$ in the clouds and below the cloud layers is consistent with tentative observations*

$NH_3$ is a necessary input for our photochemistry model, indeed the input of $NH_3$ is the core assumption of our hypothesis. We therefore discuss the tentative observations of $NH_3$ on Venus.

The Venera 8 descent probe reported the presence of $NH_3$ in the lower atmosphere of Venus. The estimated amounts from the signal are large and varied from 0.01% to 0.1%. (For further discussion on the validity of the Venera 8 $NH_3$ detection see SI Section 6). A recent re-assessment of the Pioneer Venus Large Probe Neutral Mass Spectrometer (LNMS) has also provided suggestive evidence for the presence of $NH_3$ and its oxidation products in gas phase in the cloud decks of Venus (27).

The Venera 8 observations were largely discounted at the time because $NH_3$ is not likely to be present if Venus' atmosphere is in thermodynamic equilibrium (36). At least one author supported the plausibility of the presence of $NH_3$ in the cloud layers: Florensky et al. (50) in the late 1970s argued that the upper parts of the Venus troposphere do not necessarily have to be in chemical equilibrium and could contain a number of minor chemical species, including $NH_3$ (45).

An additional argument against the plausibility of $NH_3$ is that an atmosphere containing sulfuric acid droplets cannot contain a significant amount of a free base; all of the base, in this case $NH_3$, would be sequestered in the droplets as ammonium ions. However, if the clouds have a pH > 0 and contain significant ammonium salts, then partial pressures of > 1 ppm of free ammonia gas is expected over those droplets in the lower clouds (see Methods and the SI Section 5).

Our model provides a mechanism for the release of $NH_3$ below the clouds. As the droplets gravitationally settle out of the atmosphere to higher temperatures, the droplet evaporates and $NH_3$ is released through the thermal decomposition of ammonium sulfate and ammonium sulfite. $NH_3$ is subsequently oxidized to $NO_x$ and $N_2$ (Figure 1). We note that a $NO_x$ signal has been identified in the Pioneer Venus LNMS re-analyzed data (27).

*Mode 3 Cloud Particles*

Measurements by the Pioneer Venus and Venera Probes indicate that the Mode 3 particles might not be spherical, and that their composition differs from pure concentrated sulfuric acid. (See SI Section 3 for a brief discussion of the observational support for non-spherical particles).

If $NH_3$ is the main neutralizing agent of the sulfuric acid cloud droplets, then the Mode 3 cloud particles in the lower clouds must be super-saturated in ammonium salts, with a small liquid phase, and therefore are not liquid droplets of concentrated sulfuric acid. Thus, the mechanism proposed here predicts that the Mode 3 particles in the lower cloud are solid or semi-solid, and hence likely to be non-spherical.

Specifically, the Mode 3 (largest) cloud particles in the lower cloud must be 9.3 - 18.1 molar in ammonium salts in order to provide sufficient downwards transport of $SO_2$ to produce the observed drop in $SO_2$ concentration across the clouds (see Methods and the SI Section 5). Such concentrations are not implausible if the Mode 3 particles in the cloud are actually a semi-solid slurry of ammonium salts and sulfuric acid.

We note that presence of $NH_3$ creating non-spherical Mode 3 particles is consistent with the Mode 1 and/or Mode 2 particles being of quite different composition from the Mode 3 particles. If $NH_3$ production were the result of biological activity, then life could be confined to the larger Mode 3 particles, which have more volume. If $NH_3$ was produced by a non-biological process, then it would be expected to apply to particles of all sizes, and not discriminate in favor of Mode 3 particles. However, the data on particle size and shape is consistent with Mode 1 and 2 particles being spherical (29).

Our model also explains the presence of the so-called stagnant haze layer below the cloud decks (30 km - 47 km altitude) (9). If the large Mode 3 particles are made of mostly solid ammonium sulfite and ammonium sulfate, then evaporation of any residual $H_2SO_4$ at the cloud base leaves dry solid particles. The subsequent thermal disproportionation of the remaining salts generates gas that shatters the particles at the cloud base (~100 °C at ~47 km), the fragmented particles form the haze. The haze that settles down and is not mixed back up into the clouds decomposes at ~200 °C at the bottom of the stagnant haze layer at ~30 km (Figure 1). The layered structure and the altitudes of the boundaries between the layers is therefore a natural consequence of the ammonia-based cloud chemistry. See also (7) for a discussion of the composition of the haze layer.

*$H_2S$ below the clouds*

We also note that our model predicts the presence of $H_2S$ below the clouds (Figures 1 and 3). The presence of $H_2S$ is consistent with the tentative detection of $H_2S$ below the clouds by the Venera-14 GC (24), which is the only *in situ* measured abundance value for $H_2S$. If $NH_3$ is present in the Venus atmosphere $H_2S$ is a result of disproportionation of $NH_4HSO_3$ that yields $NH_3$, $H_2S$ and $H_2O$ to the atmosphere below the clouds, and hence is a unique output of our model. $H_2S$ was also tentatively identified in the recent re-analysis of the Pioneer Venus LNMS data (27). $H_2S$, however, is a known volcanic gas on Earth so it is likely produced by volcanoes on Venus as well.

**Discussion**

Our model provides a new view of the habitability of Venusian clouds. Concentrated sulfuric acid would make the Venusian cloud environment both chemically aggressive and extremely dry (7, 14). Our model removes the issue of extreme acidity for a subset of cloud particles from consideration.

Our model implies that the Mode 3 cloud particles cannot be all composed of concentrated $H_2SO_4$. Instead, there has to be a population of cloud particles that are less acidic and have a higher pH (between -1 and 1) than concentrated sulfuric acid. Specifically, our model predicts that the Mode 3 cloud particles are semi-solid ammonium sulfites and sulfates (Figure 1) with a pH as high as 1 (see Figure 2). We emphasize that not all droplets need to contain semi-solid ammonium sulfite and (if the $NH_3$ is made by life) ammonia-producing microorganisms.

Relevant to the Mode 3 cloud particles is a recent, independent finding that the Mode 3 cloud particle composition is not primarily sulfuric acid, but instead is consistent with some particles being ammonium hydrogen sulfate ($NH_4HSO_4$), as also predicted in our analysis. Mogul et al. (58) base this finding on a re-analysis of the Pioneer Venus legacy data on the refractive index of the Venusian cloud droplets, independent of atmospheric chemistry. Also independently from our work presented here, Mogul et al. (58) have described the potential for phototropic synthesis of $NH_3$ to neutralize sulfuric acid cloud droplets, leading to the Mode 3 particle possibly containing $NH_4HSO_4$.

A pH of 0 to 1 is within the range of environments known from Earth to be habitable and in fact to be inhabited. Life can grow in acid (pH = 0) aqueous environments (35) and microbial growth in solutions as acidic as a pH = -0.5 has been claimed (59). Furthermore, most of the Mode 3 particles have been detected at altitudes in the temperature range (60-80 °C), a range that overlaps with environments known to harbor thermophilic acidophiles on Earth (with life that can grow in temperatures up to 100 °C (e.g., (60–63)).

Remarkably, examples of life on Earth secreting $NH_3$ to neutralize a droplet-sized acidic environment exist. Pathogens such as *Mycobacterium tuberculosis* and *Candida albicans* can neutralize the interior of phagosomes (acid-containing vesicles inside cells used for digestion of captured organic material), by secreting ammonia, thus evading destruction (64–66). Some plant pathogens also secrete ammonia to neutralize local pH in their target plant cells (67). By contrast, pond-dwelling acidophilic microorganisms adapt to low pH in other ways, because it is implausible for them to neutralize an entire river or pond.

Challenges to life in the Venus atmosphere remain. The extreme aridity of the Venus cloud environment has been well-known for decades (e.g. (68)), having been often described (e.g. (7, 14, 34) and most recently reviewed in (69)), and remains a significant challenge to life as we know it. Our model predicts a water vapor abundance mixing ratio of $10^{-5}$ in the lower clouds, i.e. a relative humidity of 0.02% (depending on temperature). This is ~50-fold lower than the lowest water activity known to support life on Earth. (We note that terrestrial life can survive extremely hot and dry environments as spores or other inactive forms, as summarized in the legend to Figure 1 and SI Section 10, but these are not actively growing, and to survive an ecosystem requires at least some cells or organisms to be actively growing.) The range of in-cloud water vapor abundance mixing ratios reported in the literature is very large (5 ppm to 0.2%), as summarized by (20), which may represent the presence of more clement local conditions. All global models may therefore represent an average of extremely arid 'desert' regions and much more humid 'habitable' regions.

The extreme aridity is a reflection of the very low number density of hydrogen atoms in the Venusian atmosphere. The scarcity of H atoms argues against the presence of life. Terrestrial biochemicals are typically ~50% hydrogen by atom number (as illustrated by the database of natural products compiled by (70); see SI Section 9; Table S5). However, much of the water in a bacterial cell is derived from reactions of the metabolites within the cell (71–73). For example, under active growth of *Escherichia coli*, up to 70% of the intracellular water is generated during metabolism and not transported across the membrane from the outside environment (71). If there is life on Venus, it is therefore likely to have substantially different biochemistry from Earth's, and if it is based on water as a solvent, it is likely to have very different strategies for water accumulation and retention to combat extreme aridity of the clouds. We note however that the lack of hydrogen is not just a challenge for the habitability of Venus' the clouds, but also a challenge for making detectable amounts of any hydrogen-saturated gas-phase species, such as $NH_3$, by any mechanism, abiotic or biological.

We note that additional challenges such as nutrient scarcity or high energy requirements are comparatively less limiting than aridity; for an in depth discussion of the challenges to life in the Venusian clouds see (7).

An origin for life on Venus is an open question. If life exists in the Venus clouds, it may have originated on the Venus surface, and migrated into the clouds. One model of Venus' evolution to its modern state suggests that Venus had clement surface conditions after formation, and only to have entered the current greenhouse runaway after up to 3.5 billion years (74, 75). This model is dependent on a range of specific conditions, but if correct suggests that Venus in the past had similar conditions as those under which life originated on Earth. If life emerged on the surface, terrestrial precedent suggests that some organisms would adapt to living some of the time in the clouds (reviewed in (7)). The microbial acid-neutralizing strategy provides a facile evolutionary path to Venusian cloud life. As the Venus surface became increasingly hot and uninhabitable, cloud-dwelling would become a permanent lifestyle.

As the atmospheric chemistry changed to high acidity, the cloud dwelling organisms would adapt by neutralizing their droplet habitats. A plausible evolutionary path is therefore suggested by the unique role of a droplet environment in the acid-neutralizing strategy, and the proposed history of Venus. We note however that if life is the source of $NH_3$ on Venus, it very likely does not resemble the elemental ratios of life on Earth and likely has a different biochemistry than life on Earth, specifically adapted to the unique challenges of the Venusian cloud environment.

The Venus low D/H ratio (76) and the possible existence of felsic rocks which form in the presence of water (77–80) imply the presence of past Venus oceans, yet the debate on whether or not Venus ever had oceans continues. Recently, Turbet et al. demonstrated with a 3D global climate model that Venus may have been too hot early on for water oceans to form (81). Their climate model shows that the steam atmosphere of early Venus' never condensed on the planet's surface to form liquid water oceans. Instead, accordingly to the model, water vapor condensed on the nightside of the planet to form clouds that warmed the surface by absorbing and re-emitting the planet's outgoing infrared radiation (81). However, Turbet et al. do state that a comprehensive sensitivity study is needed to quantitatively confirm their result, as cloud and atmospheric circulation feedbacks can vary nonlinearly and non-monotonically with rotation period (81). The newly selected VERITAS and EnVision missions, as well as DAVINCI's should solidify or rule out the possibility of the past water-rich era of Venus, by a combination of D/H measurements and multispectral imaging of the tesserae regions for mineral compositions.

**Summary and Critical Future Measurements**

Our hypothesis of locally produced $NH_3$ in the Venus clouds explains a number of anomalies in the atmosphere and clouds of Venus. Our photochemical model of the consequences of $NH_3$ production explains: the $SO_2$ depletion in the clouds and vertical abundance profile of $H_2O$, building on the work of (20); the presence of $O_2$ in the clouds; supports the in situ detection of $H_2S$ below the clouds, and explains the non-spherical nature of Mode 3 particles. While the presence of other mineral bases could contribute, none of them can explain the ppm levels of $O_2$ in the clouds nor the tentative presence of $NH_3$. No definitive source for $NH_3$ has been identified; in chemical terms, biological production is the most plausible, but the concept of life in the clouds of Venus remains controversial. Many of the *in situ* observations should be repeated for confirmation and more model work is needed to fully resolve the vertical abundance profiles of relevant gases.

We must be careful not to fall for a conjunction fallacy. While life may explain the combined anomalies with some external assumptions, there may as yet be a chemical explanation for each individual anomaly.

An *in situ* Venus probe can support or refute our proposed view of Venus as an inhabited planet with the following measurements.

Gases
- Establish the existence of $NH_3$ and $O_2$ in the cloud layers.
- Measure the amounts of $NO_x$ to establish which $NH_3$-destruction pathway dominates.
- Determine the specific altitude-dependent abundance profiles of $H_2O$, $SO_2$, and $H_2S$, ideally with day and night measurements to inform chemistry sources and sinks.

Cloud particles
- Confirm the non-spherical, semi-solid nature of Mode 3 cloud particles and identify them as ammonia salts.
- Measure the pH of cloud particles, especially Mode 3 cloud particles
- Detect organic molecules in cloud particles; if found exclusively in the larger particles would be an indicator of life.

Search for life
- Analyze a large number of individual cloud particles, especially Mode 3, for morphological and chemical signs of life.

In the meantime, a public release of original data from the Russian Venera and Vega missions could enable further support or refute of current models and predictions, and would provide needed context for future mission results.

We have presented an initial analysis of several sources for the $NH_3$ on Venus. We have argued that biological production may be a potential source of both $NH_3$ and $O_2$ that we have identified that meets the quantitative requirements for $NH_3$ production. Although the biomass required to make $NH_3$ and $O_2$ at the required rate is not unrealistic, at 0.05% the total biomass on Earth and ~1.5% of the total Venusian cloud mass, life in the clouds of Venus has been considered implausible because of very high acidity, very low water activity, and scarcity of hydrogen atoms. By predicting a Mode 3 particle pH of -1 to 1 due to neutralizing $NH_3$, our work implies both that Venus clouds are more habitable than previously thought, and, by the requirement of locally-produced $NH_3$, that clouds may be inhabited. We hope our work will encourage further studies into habitability and astrobiological potential of Venusian clouds.

**Materials and Methods**

**Photochemical model.** The details of the model are provided in the SI, Section 4. In summary, we employ a 1D Lagrangian photochemistry/diffusion code that follows a single parcel as it moves from the bottom to the top of the atmosphere. The temperature, pressure, and actinic ultraviolet flux are prescribed at each altitude in the atmosphere (20).

**Calculation of flux of ammonia.** We calculate the flux of ammonia necessary to maintain the observed gradient of $SO_2$ through the clouds following the method of Rimmer et al (20). The goal is to explain the removal of nearly all of $3.5 \cdot 10^{15}$ cm$^{-3}$ of $SO_2$ ($1.5 \cdot 10^{-4}$ bar at 300 K level of the atmosphere) that should be present from upward mixing from volcanic sources and recycled $SO_2$.

The time taken for $SO_2$ to mix through the region 45 – 65 km is calculated using Lee et al. 2015, their Eq. 7, (82):

$$\text{Time} \approx 2 \cdot \frac{H^0 \cdot \delta h}{K_{zz}} = 2.6 \cdot 10^8 \text{ seconds} \approx 8.25 \text{ Earth years}.$$

In other words, $SO_2$ will be replenished in the atmospheric cloud layers in 8.25 Earth years, and this is the timescale that the presence of $NH_3$ needs to remove $SO_2$. The atmospheric scale height $H_0 \approx 6.5 \times 10^5$ cm is the average scale height in the atmospheric cloud layers, $\delta h = 2 \times 10^6$ cm is the distance between 45 and 65 km, and $K_{zz} = 10^4$ cm$^2$ s$^{-1}$ is the eddy diffusion coefficient throughout the atmospheric cloud layers. The flux (cm$^{-2}$ s$^{-1}$) of $SO_2$ into the clouds is therefore given by

$$\Phi = \frac{\text{Amount} \cdot \text{Distance}}{\text{Time}} = \frac{(3.5 \cdot 10^{15} \text{cm}^{-3}) \cdot (2 \cdot 10^6 \text{ cm})}{2.6 \cdot 10^8 \text{ sec}} = 2.7 \times 10^{13} \text{ cm}^{-2} \text{ sec}^{-1}.$$

Recall there is a one-to-one molar ratio for $NH_3$ to remove $SO_2$. Given the mass of $NH_3$, is the flux rate above is equivalent of $1.1 \times 10^{11}$ tonnes per year of $NH_3$.

**Calculation of concentration of $NH_3$ in particles.** The concentration of salts in the cloud droplets can be estimated from the concentration necessary to provide the flux of $NH_3$ as calculated above. The necessary flux of $NH_3$ is dependent on the size of the particles, and hence the particles' rate of fall. For a given particle size we can calculate the rate of fall, and hence the volume of cloud material removed per second, and from this the concentration of salts in that volume needed to provide the flux calculated above. See SI, Section 2 for the details on the calculation of the concentration of ammonium salts in the lower cloud particles

**Calculation of concentration of gaseous $NH_3$ over droplets.** The concentration of gaseous $NH_3$ over an acid droplet containing dissolved $NH_4^+$ was calculated as follows. The fraction of total N species that is present as $NH_3$ and as $NH_4^+$ can be calculated from the pKa of $NH_3$ as the pH. The concentration of $NH_3$ over solution can be calculated from the solution concentration and Henry's constant (see SI, Section 5, Figure S3). Both pKa and Henry's constant are dependent on temperature.

### Acknowledgments

P.B.R. thanks the Simons Foundation for funding (SCOL awards 599634). S.S. thanks the Change Happens Foundation and by Breakthrough Initiatives for partial funding of this work. We are grateful to Vladimir Krasnopolsky for useful discussions on the presence of $O_2$ in the atmosphere of Venus.

**Figures and Tables**

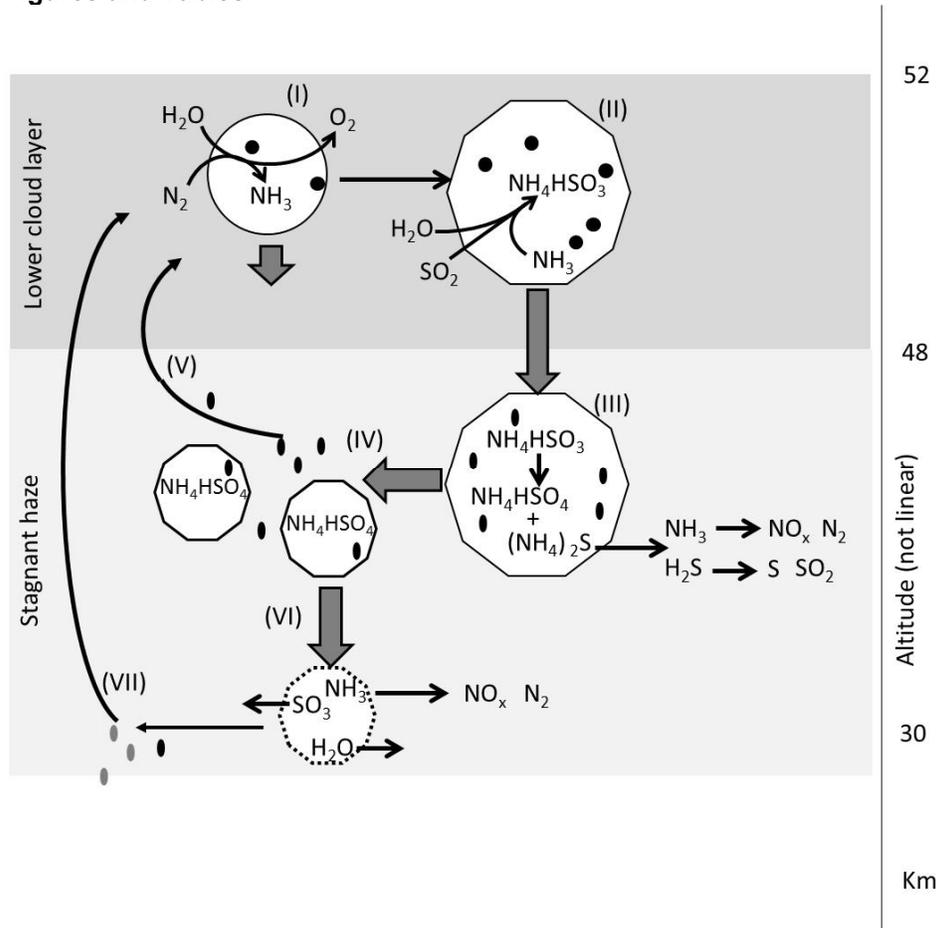

Figure 1. Ammonia cycle in the atmosphere of Venus. See SI Section 10 for details. (I) $NH_3$ is produced locally in the clouds from atmospheric $N_2$ and $H_2O$ (Table 1) by metabolically active microorganisms (black dots) inhabiting cloud droplets (white circle). (II) The production of $NH_3$ in the droplet raises the droplet pH to -1 to 1 (from -11 on the Hammett acidity scale) by trapping the $SO_2$ and $H_2O$ in the droplet as ammonium hydrogen sulfite ($NH_4HSO_3$). The production of sulfite salts in the droplet leads to the formation of a large, semi-solid (and hence non-spherical) Mode 3 particle (white decagon). (III) The Mode 3 particle settles out of the clouds where ammonium sulfite disproportionates to ammonium sulfate and ammonium sulfide; the latter decomposes to $H_2S$ and $NH_3$, which in turn undergo photochemical reactions to a variety of products. (V) Disproportionation and gas release break up the Mode 3 particles into smaller haze particles and microorganism spores (black ovals), some of which return to the cloud layer (V). (VI) The ammonium sulfate particles fall further below the cloud decks, where ammonium sulfate decomposes to $SO_3$, $NH_3$ and $H_2O$. (VII) Spores released at this stage may be unviable (grey ovals), but any surviving could also be eventually transported back to the clouds.

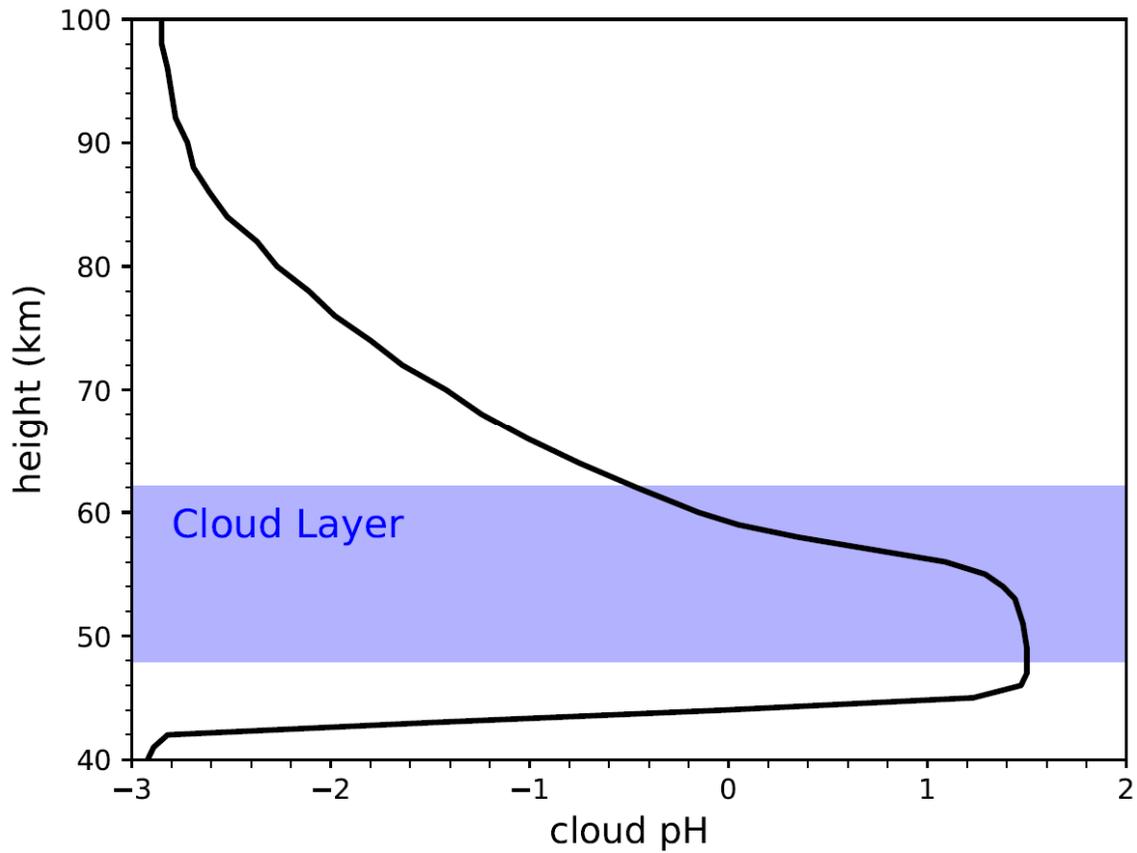

Figure 2. Predicted pH profile of cloud particles. The blue shaded region shows the altitude where clouds are present, from 48 – 62 km. Note that the plot extends above and below the cloud tops because there are plausibly cloud particles populations that extend down to the altitude where sulfuric acid is sublimated, and up into the mesosphere where sulfuric acid aerosol evaporation may explain the anomalous $SO_2$ inversion at 80-100 km. Our model provides no constraints on the composition of the mesospheric particles, which may well be composed of pure sulfuric acid.

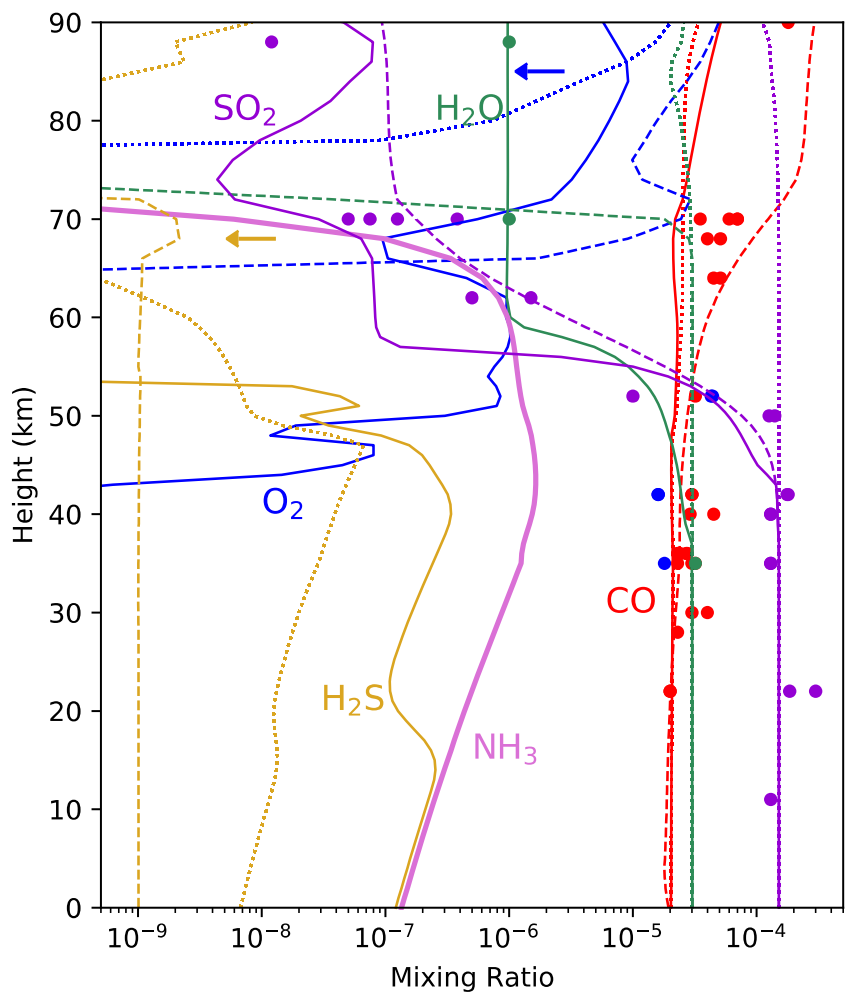

Figure 3. Venus atmosphere abundance profiles of key molecular species. The *x*-axis is the gas fraction by volume, called the mixing ratio. The *y*-axis is altitude above the surface in km. The lines are gas mixing ratios from our models: with $NH_3$ chemistry (solid lines); without $NH_3$ chemistry (dotted line; model in (20)); without $NH_3$ but with an arbitrary removal rate for $SO_2$ in the cloud layers tuned to fit the data (dashed lines; model in (20, 34)). The colored circles show a representative subset of collated remote and *in situ* data (error bars not shown) from (20) (their Table 4) and (33) (their Supplementary Table 3). Key is that the baseline model predicts no $NH_3$ or $H_2S$ above the 1 ppb level. Models with $NH_3$ chemistry have very different $H_2O$, $SO_2$, $O_2$, and $H_2S$, values at some altitudes than models without $NH_3$ chemistry, and improve the match to observational data. The main takeaway is that the model without $NH_3$ and without the $SO_2$ arbitrary removal rate (dotted line) fit the cloud-layer data very poorly whereas the model with $NH_3$ (with no arbitrary constraints; solid line) fits the data much better. The boundary conditions for surface abundance in the photochemical model are listed in Table S6 in the SI.

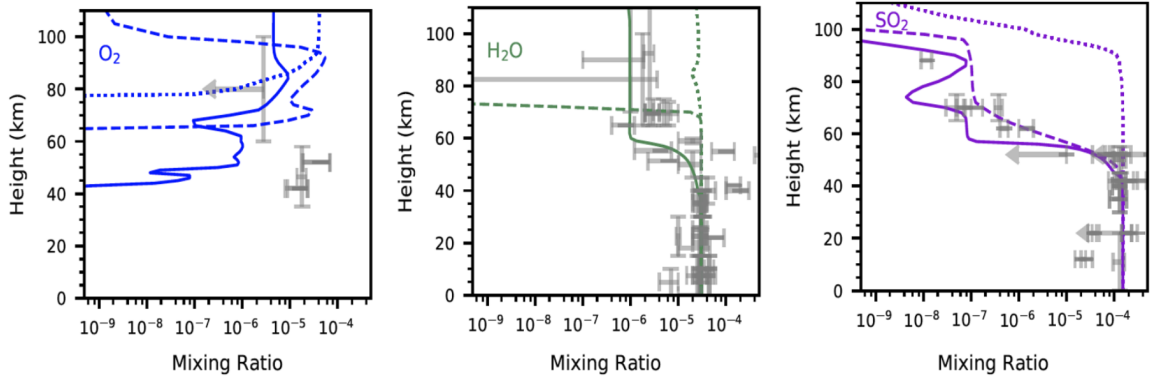

Figure 4. Venus atmosphere abundance profiles of three molecular species. The *x*-axis is the gas fraction by volume, called the mixing ratio. The *y*-axis is altitude above the surface in km. The lines are gas mixing ratios from our models: with $NH_3$ chemistry (solid lines); without $NH_3$ chemistry (dotted line; model in (32)); without $NH_3$ but with an arbitrary removal rate for $SO_2$ in the cloud layers tuned to fit the data (dashed lines; model in (20, 34)). Gray points with error bars are data from observations tabulated in (20). Left: $O_2$. Our model with $NH_3$ chemistry improves upon both the long-standing problem of presence and overabundance of $O_2$ in the upper atmosphere and the presence of $O_2$ in the cloud layers. Middle: $H_2O$. Our model with $NH_3$ chemistry supports the presence of water vapor above the cloud layer (> 80 km). Right: $SO_2$. Our models with $NH_3$ chemistry (solid line) and without $NH_3$ chemistry but with arbitrary constraints on $SO_2$ (dashed line) both provide a fit to observed values throughout the atmosphere except for the top (> 85 km). Key is that the model without $NH_3$ and without the $SO_2$ arbitrary removal rate (dotted line) fit the cloud-layer data very poorly whereas the model with $NH_3$ (with no arbitrary constraints; solid line) fits the data much better.

Table 1. Free energy per mole for $NH_3$-generating reactions under Venus cloud conditions.

| Reaction | Free energy of reaction (kJ/mol) | Free energy required per mole of surplus $NH_3$ (kJ/mol) | Water consumed per surplus $NH_3$ |
|---|---|---|---|
| 1  $4N_{2(aq)} + 11H_2O_{(l)} \rightarrow 2NH_4^+OH^-_{(aq)} + 3NH_4^+NO_3^-_{(aq)}$ | 1730 – 2024 | 865 – 1012 | 6.5 |
| 2  $N_{2(aq)} + 8H_2O_{(l)} \rightarrow 2NH_4^+OH^-_{(aq)} + 3H_2O_{2(aq)}$ | 1203 – 1471 | 602 – 736 | 4 |
| 3  **$2N_{2(aq)} + 10H_2O_{(l)} \rightarrow 4NH_4^+OH^-_{(aq)} + 3O_{2(aq)}$** | **1000 – 1306** | **262 – 343** | **2.5** |
| 4  $4N_{2(aq)} + 17H_2O_{(l)} + 3HCl_{(aq)} \rightarrow 5NH_4^+OH^-_{(aq)} + 3NH_4^+ClO_4^-_{(aq)}$ | 1364 – 1634 | 273 – 323 | 3.4 |
| 5  $N_{2(aq)} + 6H_2O_{(l)} + 3SO_{2(aq)} \rightarrow (NH_4^+)_2SO_4^{2-}_{(aq)} + 2H_2SO_{4(aq)}$ | 1193 – 1313 | N/A | N/A |

Free energies of $NH_3$-producing reactions calculated from (83–85). Ranges are minimum to maximum over a range of pH = -3 to pH = +4 and temperature from 2 °C to 115 °C. Concentrations of $SO_2$ and $H_2O$ are as described in (34). $O_2$ fractional abundance is assumed to be $10^{-6}$. Table columns are as follows. First column: reaction number. Second column: possible chemical reaction that produces $NH_3$. Third column: free energy of reaction assuming that $NH_3$ is accumulated to 2 molar concentration. For the fourth and fifth column, values were calculated in terms of 'surplus $NH_3$', which is the amount of $NH_3$ synthesized as $NH_4OH$. Fourth column: free energy per mole of 'surplus $NH_3$' produced. Fifth column: number of water molecules consumed per 'surplus' $NH_3$. Reaction number 3 (black bold font), which produces molecular oxygen as an oxidized byproduct, is the most efficient, both in its use of energy and its use to water. We note that Reaction 4 could produce hypochlorite, chlorite, or chlorate as an oxidized product, but as perchlorate is relatively stable and is the weakest oxidizing agent, we have shown this reaction for illustration only. Reaction 5 generates more acid than it consumes, and so cannot be a source of the base which neutralizes $H_2SO_3$. We also note that Reaction 1 and Reaction 4 (reactions making nitrate and perchlorate respectively) cloud also alternatively explain the presence of $O_2$. Nitrate and perchlorate would "rain out" and decompose to $N_2$ and $O_2$ or HCl, $Cl_2$ and $O_2$ respectively below the clouds. *In situ* measurements of $NO_x$ and $ClO_4$ abundance in the clouds could rule out these reactions as a potential source of indirect formation of $O_2$.

# Supplementary Information for
Production of Ammonia Makes Venusian Clouds Habitable and Explains Observed Cloud-Level Chemical Anomalies

William Bains[1,2], Janusz J. Petkowski[1], Paul B. Rimmer[3,4,5], Sara Seager[1,6,7,*]

[1]Department of Earth, Atmospheric, and Planetary Sciences, Massachusetts Institute of Technology, Cambridge, MA 02139, USA
[2]School of Physics & Astronomy, Cardiff University, 4 The Parade, Cardiff CF24 3AA, UK
[3]Department of Earth Sciences, University of Cambridge, Downing St, Cambridge CB2 3EQ, United Kingdom
[4]Cavendish Laboratory, University of Cambridge, JJ Thomson Ave, Cambridge CB3 0HE, United Kingdom
[5]MRC Laboratory of Molecular Biology, Francis Crick Ave, Cambridge CB2 0QH, United Kingdom
[6]Department of Physics, Massachusetts Institute of Technology, Cambridge, MA 02139, USA
[7]Department of Aeronautics and Astronautics, Massachusetts Institute of Technology, Cambridge, MA 02139, USA.

*Sara Seager

**Email:** seager@mit.edu

**Supplementary Information Text**

**1. Ammonia is the only plausible base that can be made locally in clouds.**

A base is required that can play the role of B in the reaction:

$SO_2 + H_2O + B \rightarrow SO_2 + BH^+ + OH^- \rightarrow BH^+ + HSO_3^-$.

The pKa of the buffer $H_2SO_3/HSO_3^-$ is 1.71, so we require a base for which the pKa of the pair B + $H^+$ / $BH^+$ is greater than 1.71, to that there is a pH ($H^+$ concentration) at which $HSO_3^-$ is favored over $H_2SO_3$ and $BH^+$ is favored over B. If the base is to be formed locally in the clouds and not transported from the surface, then it must be formed from locally available elements. Of the possible bases that could be formed from the elements present in volatile compounds in the clouds (H, C, N, O, F, P, S, Cl), only ammonia ($NH_3$) or hydrazine fulfil that criterion as described in Table S1.

**2. Calculation of $NH_4^+$ salt concentration in cloud droplets.**

The rate of settling of cloud particles can be estimated as follows. The lower cloud is characterized by particles in three size modes (1, 2). These particles are assumed to be roughly spherical. Assuming that they are small enough to fall under a laminar flow (Stokes) regime, then their settling velocity can be calculated from

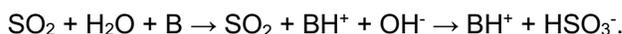

$v = \frac{2}{9} \cdot \frac{\rho}{\mu} \cdot g \cdot r^2 . 100$   cm/sec,      (1)

where $v$ = terminal velocity in cm/sec, $\rho$ = droplet density in kg/m³ (assuming that this is very much greater than the gas density, so the gas density can be neglected), $\mu$ = dynamic viscosity of the gas, g = acceleration due to gravity in m/sec², which on Venus is 8.87 m/sec², and $r$ is the particle radius in meters. μ for pure $CO_2$ at 100 °C is ~1.85·10⁻⁵ kg/m/sec, and this is taken as the viscosity of the Venusian atmosphere. The particles are assumed to be roughly spherical – a slurry of liquid and solid is likely to adopt a roughly spherical form in free fall. The density of ammonium hydrogen sulfate is 1.79 g/cm³, the density of sulfuric acid is 1.74 -1.82 g/cm³, depending on concentration. We therefore adopt a density of 1.75 g/cm³.

The rate at which material falls from the clouds is therefore

$m_f = d \cdot v$  g/cm²/sec,                    (2)

where $m_f$ is the mass falling past a 'cloud base' level, $d$ is the mass density of cloud particles and $v$ is the velocity. The mass density depends on the density of the particles, which itself depends on the concentration of solutes in $H_2SO_4$ (assuming that $H_2SO_4$ is the solvent phase). At the cloud base the droplets have previously been assumed to be ~100% sulfuric acid at ~100 °C, which has a density of 1752 kg/m³ (3). There is no data on the density of ammonium sulfite solutions in concentrated sulfuric acid, but the density of ammonium sulfate in concentrated sulfuric acid is only slightly different from the density of acid (4).

The size distribution of the droplets is taken from (2) (although we note that their modelling of the clouds has been contested (5, 6)). The distribution of cloud particles was taken from the analysis of the Pioneer Venus sounder measurements reported in (2), for the 49 km (i.e. lower cloud) layer. They deduced a trimodal cloud distribution, with the distribution as described in Table S2.

Knollenberg and Hunten (ref. (2)) note that the log normal distribution of Mode 3 particles significantly under-estimates the abundance of the largest Mode 3 particles. We confirm this in Figure S1, top panel, where distributions have been fitted to the observed abundance data for the 49 km sample, from Figure 12 of (2). Large particles have a disproportionate influence on the rate of loss of material from the cloud by setting for two reasons; they have larger volume and hence contain more mass, and have a larger cross-sectional area and hence settle faster. We therefore adopted a log-log-normal distribution (i.e. a distribution that is normal for log(log(diameter))), a function that is not defined for diameter < 1) for Mode 3 particles, which fits the observed data better (Figure S1, bottom panel). We do not claim that this has a realistic physical interpretation, just that this interpolates the observed particle distribution better than a log-normal distribution.

This particle distribution is expected to settle at a rate yielding $6.71 \cdot 10^{-12}$ liters/cm$^2$/second. 97.6% of this settling is carried by Mode 3 particles. The value of the necessary flux of SO$_2$ from the clouds from equation (1) can now be compared to the flux of particle mass from the clouds in equation (2). Assuming a density of 1750 kg/m$^2$ for the droplets made mostly of sulfuric acid or its salts, the concentration of SO$_2$ is

$$[SO_2] = \frac{\rho}{N_a} \cdot \frac{d}{m_f} = 9.34 - 18.05 \text{ molar} \qquad (3)$$

where $\rho$ and $m_f$ are as defined above, $d$ is the density of the cloud droplets and $N_a$ is Avagadro's number, $6.023 \cdot 10^{23}$, with a range of values depending on whether the mean Mode 3 particle diameter is 7.5 or 8.5 microns. A saturated solution of ammonium hydrogen sulfite in water is ~5 molar. The density of solid ammonium sulfite is 1.41g/cm$^3$ (7) (no density of available for solid ammonium hydrogen sulfite), which implies that solid ammonium sulfite has ~12 moles per 1000 cm$^3$. Thus, for settling of cloud particles to be the sole removal mechanism for ammonium sulfite, and ammonium sulfite formation to be the sole mechanism for removal of SO$_2$, the cloud particles must be super-saturated in ammonium salts, and may be mostly solid ammonium salts with a small liquid phase of aqueous sulfuric acid. We note that this model presents the "upper limit" of ammonium sulfite concentration needed to explain the depletion of SO$_2$ in the clouds, as NH$_3$ does not have to be the only neutralizing agent in the clouds (8).

**3. Evidence that Mode 3 particles are non-spherical.**

Estimates of the refractive index of the particles in the lower cloud suggest a value of ~1.33 assuming spherical droplets (9), which is lower than any plausible value for concentrated sulfuric acid. The droplets could be more dilute sulfuric acid, but this is not compatible with the vapor pressure in the clouds (10). This anomaly can be resolved if the particles absorb a small amount of incident light; however (11) find no noticeable absorption in the lower clouds. An alternative explanation is non-spherical particles (2, 9). which implies non-liquid ones. The Pioneer Venus sounder data supports non-spherical particles for the largest Mode 3 particles. The optical array spectrometer (OAS) instrument had three photodiode arrays which measured the shadow of particles as they passed, which makes the particle size measurement independent of particle composition. In the lower cloud, the range 2 array (which measures particles of 5 – 53 μm diameter) counted 72 particles larger than 16 μm, whereas the range 3 array (which measures particles of 16 – 181 μm diameter) counted only 3 particles (2). The comparison is complicated by different sampling volume and different sensitivities, but a discrepancy significant to p<0.01 remains. Such discrepancies are well-known in using OAS instruments to measure non-spherical particles such as snowflakes (12). The particle size distribution as determined by the OAS and reflectance as determined by the solar flux radiometer are also inconsistent with spherical particles (13).

We note that the early Venera and Vega measurements of the cloud particles' properties are consistent with data acquired by Pioneer Venus (14) and likewise have also returned inconsistent data about the cloud composition. We acknowledge however that the Mode distribution of the Venus' cloud particles is a topic of decades-long heated debate. Several studies questioned the existence of the large Mode 3 particles altogether, e.g. (6, 15), and for example claimed that Mode 3 could in fact be a large "tail" of the liquid Mode 2' distribution, once calibration errors are taken into account (5).

Such decades-long lingering questions on the true nature of the Venus cloud particles strengthen the need for a renewed campaign of in situ measurements to characterize the aerosols.

**4. Details of atmospheric photochemistry model.**

In the rest-frame of the parcel, diffusion terms are accounted for by time-dependence of the chemical production, $P_i$ (cm³ s⁻¹), and loss, $L_i$ (s⁻¹), and so below the homopause, resulting in the equation:

$$\frac{\partial n_i}{\partial t} = P_i[t(z, v_v)] - L_i[t(z, v_z)]n_i, \qquad (4)$$

where $n_i$ (cm⁻³) is the number density of species $i$, $t$ (s) is time, $z$ [cm] is atmospheric height. Convergence conditions and further details, including the chemical network, are given in (8). We implement the effective rates for SO₂ depletion into the clouds under the assumption that the NH₃ chemistry is sufficient to regulate the in-cloud SO₂ abundances. We modify the model to add in-cloud fluxes of O₂, NH₃, H₂S and SO₃. The degassing rates represent the following reactions:

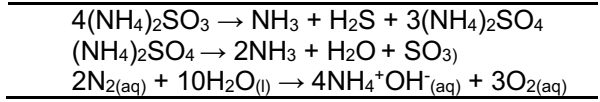

$$4(NH_4)_2SO_3 \rightarrow NH_3 + H_2S + 3(NH_4)_2SO_4$$
$$(NH_4)_2SO_4 \rightarrow 2NH_3 + H_2O + SO_{3)}$$
$$2N_{2(aq)} + 10H_2O_{(l)} \rightarrow 4NH_4^+OH^-_{(aq)} + 3O_{2(aq)}$$

with the rates defined as:

Eq. (5):
$$\Phi(O_2) = \Phi_0 \left\{ \frac{1}{2}\tanh\left[\frac{1}{2}(z - h_4)\right] \cdot \tanh\left[\frac{1}{2}(h_6 - z)\right] + \frac{1}{2} \right\}$$

Eq. (6):
$$\Phi(H_2S) = \frac{1}{2}\Phi_0 \left\{ \frac{1}{2}\tanh\left[\frac{1}{2}(z - h_3)\right] \cdot \tanh\left[\frac{1}{2}(h_5 - z)\right] + \frac{1}{2} \right\}$$

Eq. (7):
$$\Phi(SO_3) = 3\Phi_0 \left\{ \frac{1}{2}\tanh\left[\frac{1}{2}(z - h_1)\right] \cdot \tanh\left[\frac{1}{2}(h_2 - z)\right] + \frac{1}{2} \right\}$$

Eq. (8):
$$\Phi(NH_3) = \Phi_0 \left\{ \frac{1}{2}\tanh\left[\frac{1}{2}(z - h_3)\right] \cdot \tanh\left[\frac{1}{2}(h_5 - z)\right] + \frac{1}{2} \right\} + 3\Phi_0 \left\{ \frac{1}{2}\tanh\left[\frac{1}{2}(z - h_1)\right] \cdot \tanh\left[\frac{1}{2}(h2 - z)\right] + \frac{1}{2} \right\}$$

where $h_1$ = 30 km, $h_2$ = 35 km $h_3$ = 40 km, $h_4$ = 45 km, $h_5$ = 50 km and $h_6$ = 65 km. The heights are set based on altitudes i) at which ammonium sulfate decomposes ($h_1$ and $h_2$), ii) at which ammonium sulfite disproportionates (between $h_3$ and $h_5$), and iii) at which oxygen is presumed to be produced in the lower cloud (between $h_4$ and $h_6$). The altitudes $h_3$, $h_4$, $h_5$ and $h_6$ are set by the temperature profile of Venus's atmosphere. $\Phi_0$ = 1.5 ·10⁷ cm⁻³ s⁻¹ is the degassing rate needed to account for the amount of ammonium sulfite sufficient to store the in-cloud $SO_2$. The leading factors are stoichiometric. The overall fluxes of our model are shown on Figure S2.

The boundary conditions for surface abundance in the photochemical model are listed in Table S6.

### 5. Prediction of vapor pressure of $NH_3$ over solution of different pH.

The abundance of $NH_3$ in the atmosphere over a solution of given pH was calculated as follows. The abundance of free $NH_3$ in a solution of ammonium salts is given by

$$pKa = \log_{10} \cdot \frac{[NH_3] \cdot [H^+]}{[NH_4^+]},$$

where $[NH_3]$ is the concentration of free $NH_3$ in solution. The concentration of $NH_3$ in gas phase is given by

$$[NH_{3(g)}] = H \cdot [NH_{3(aq)}],$$

where *H* is Henry's constant. The pKa for $NH_3$ as a function of temperature and pH was calculated as described by (16). Henry's law as a function of temperature was taken from averaged values presented by (17). The result for different pH values the Mode 3 particles is shown in Figure S3.

### 6. Commentary on the Venera 8 detection of $NH_3$.

The Venera 8 gas analyzer had a dedicated instrument to detect $NH_3$ using bromophenol blue as an indicator of a basic atmospheric component (18). This experiment for the determination of the $NH_3$ mixing ratio in the 44 to 32 km altitude region estimated the $NH_3$ mixing ratio between 0.01% and 0.1%, from the color change of "bromophenol blue" (18). The Venera 8 values are difficult to reconcile with the proposed 6 ppb upper limits for $NH_3$ abundance above the clouds (19), unless the $NH_3$ loss in the upper atmosphere is balanced by a constant production that is localized to the lower atmospheric regions (the clouds and the stagnant haze layer below). Such discrepancies can only ultimately be resolved by new in situ measurements of $NH_3$ in the clouds of Venus.

It appears that the instrument detected absorbance change in bromophenol blue exposed to atmosphere compared to a control chamber; however absorbance was measured by photoresistor response to an (undescribed) light source (18, 20), and so might not have been able to distinguish color change due to pH change from color change resulting from indicator breakdown, as also pointed out by (21) who postulated that the Venera 8 $NH_3$ detector could have responded to gaseous sulfuric acid, as bromophenol blue turns violet-red in concentrated sulfuric acid.

### 7. Non-biological pathways of formation of $NH_3$ in the clouds of Venus.

### 7.1. Production of $NH_3$ by lightning.

We model the production of $NH_3$ by lightning following the model of (22). The molecules intercepted by a lightning bolt are assumed to be broken into their component atoms, and then these recombine at random depending solely on the relative numbers of atoms. Thus, to make $NH_3$, a nitrogen atom must collide with three hydrogen atoms and *not* collide with any other atom. Thus, the fraction of N atoms that form $NH_3$

$$f(N) = \left(\frac{H}{T}\right)^3,$$

where f(N) is the fraction of N atoms that form $NH_3$, H is the number of N atoms/$cm^3$ and T is the total number of atoms /$cm^3$. Assuming 100 lightning strikes/second, a ratio of cloud:cloud vs cloud:ground lightning strikes of 100:1, 100 diameter lightning bolts that are 10 km long in the clouds and 45 km long between clouds and ground, and using the range of gas concentrations used in (22), we find the production rate of $NH_3$ is 16.7 moles/year (standard deviation 1.26), i.e. 285.6 grams/year. An example of the calculation is given in Table S3.

The examples of the calculations presented in Table S3 assume a lightning strike rate similar to Earth. The rate would have to be ~$10^9$ times the terrestrial value to approach the production rate needed to explain the presence of $NH_3$ in the clouds, even assuming this mechanism. Such extreme lightning activity seems implausible. This is in line with previous work on $NH_3$ as a biosignature gas (23).

As an extreme calculation, we calculated what the thermodynamic equilibrium concentration of $NH_3$ and oxygen would be if the entire atmosphere was 'shocked' to high temperature and quenched at 1200 K, the temperature at which lighting or shock-driven reactions can be considered to be quenched (Zahnle et al. 2020, their Fig. 1 (24)). The result is shown in Table S4. Such a scenario would require the entire atmosphere to be filled with lightning, not dissimilar to the scenario required in the calculation above, in Table S3.

### 7.2. UV photolysis producing $NH_3$.

We can model the photochemical production of $NH_3$ in two ways. The first is to perform a full photochemical model of the atmosphere, using available kinetic data. This has been done previously, and suggests negligible (~ parts per tirillion, ppt) $NH_3$ abundance (e.g. the Rimmer et al hydroxide cloud chemistry model achieves ~11 ppt of $NH_3$ (8)).

There remains the possibility that non-equilibrium photochemistry using pathways not modelled could generate $NH_3$. Given that the photochemistry of nitrogen species in gas phase has been explored very extensively, the most likely context for this would be the photochemistry of $N_2$ in sulfuric acid solution, which to our knowledge has not been explored. $N_2$ is not protonated in $H_2SO_4$; indeed, $N_2$ is frequently used as an inert carrier gas for investigations of reactions of gases with $H_2SO_4$. It also is unlikely that reaction in $H_2SO_4$, a powerful oxidizing agent, would produce a reduced gas. However the possibility remains, and could be explored experimentally.

### 7.3. Volcanic production of $NH_3$.

For the purposes of exposition, we assume that Venusian volcanoes produce $NH_3$ at the same rate as terrestrial volcanic and hydrothermal systems. This will grossly over-estimate production of $NH_3$, as the large majority of terrestrial $NH_3$ is produced as a result of the presence of abundant water in volcanic and hydrothermal systems (for example, by reaction in serpentizing systems). However, this provides as basis for comparison.

A tabulation of the fraction of gas over 100 °C emitted by terrestrial fumaroles is summarized in Figure S4. (Gas below 100°C is liable to condense and wash out any $NH_3$ present).

The geometric mean ratio $NH_3/CO_2$ was 0.004. Estimates of the global volcanic $CO_2$ emission is 6 – 11 x $10^{12}$ moles/year (25). This suggests a flux of $NH_3$ of a maximum of ~640 g/second. This is ~5 x $10^4$ times too low a rate to fit the model presented here. For volcanism to be a plausible source of the $NH_3$ required by this model, a) volcanism on Venus would have to be more than $10^4$ times as volcanically active as Earth and b) Venusian volcanism must involve either as much water as terrestrial volcanism or have an equally abundant, alternative source of hydrogen. Neither requirements appear plausible.

We note that trace $NH_3$ can also be produced by some, but not all, volcanic systems on Earth (see (26) and their Table 2.38). For example, (27) reported an emission of volcanic $NH_3$ from the volcano on Miyake-jima island, in the Izu archipelago, that reached 5 ppb locally. The likely source of $NH_3$ in such cases is the thermal breakdown of organic matter in the crust that has accumulated in the oceanic sediments. The organic matter reacts with the rising magma beneath the volcano, producing $NH_3$ gas (27, 28). This mechanism would only apply if there were coal seams on Venus, which itself would be a strong biosignature, albeit one hard to test without drilling.

**8. Non-biological pathways of formation of molecular oxygen in the clouds of Venus.**

**8.1. Production of $O_2$ by lightning.**

The calculation above also suggests that, despite oxygen atoms being abundant in the atmosphere, the production of $O_2$ is very inefficient and will result in very low abundances of $O_2$. In addition, any lightning production of $O_2$ will destroy any > 1 ppt concentrations of $NH_3$, dissociating the molecule alongside $CO_2$. The end result will be conversion of the majority of the $NH_3$ into NO, $N_2$ and $H_2O$. Lightning production of $O_2$ is not compatible with the survival of $NH_3$.

**8.2 Production by thermal breakdown of $H_2SO_4$.**

$H_2SO_4$ is thermally dissociated into $SO_3$ and $H_2O$ below ~35 km in Venus atmosphere (29). The equilibrium amount of $O_2$ formed by the reaction

$2SO_3 \rightarrow 2SO_2 + O_2$,

was calculated as follows. The standard free energy of the reaction was calculated from NIST/JANAF tables (17). The abundance of $SO_3$ is calculated by (30) to be very low in the lower atmosphere, and was taken to be $10^{-7}$. The abundance of $SO_2$ was taken from the same source to be $2 \times 10^{-4}$. From this, the equilibrium partial pressure of $O_2$ can be calculated for a given pressure. As the pressure for temperatures >730 K (i.e. below the Venusian surface) are speculative, calculations were done at 1 bar – lower pressure will favor $O_2$ production, and so this favors $O_2$ production in the lower atmosphere. The results are plotted in Figure S5. This shows that at Venus surface temperature the expected abundance of $O_2$, if $SO_3$ is present at the relatively high abundance of $10^{-7}$, is ~$5 \times 10^{-12}$, i.e. 6 orders of magnitude less than that reported by Pioneer Venus and required by our model.

Also plotted for comparison is the thermodynamics of the industrial processing pure $H_2SO_4$, i.e. partial pressure $SO_3$ = 0.5, and 10% $SO_2$. This shows that the industrial process is indeed thermodynamically efficient at breaking down $SO_3$ at temperatures around 700 °C.

**9. Hydrogen in biological molecules.**

We analyzed three databases of biological molecules for the relative abundance of hydrogen atoms. A database of ~200,000 natural products (31) has been exhaustively validated as exclusively being bona fide products of metabolism in a diverse range of organisms. We also used the Roche Biochemical Pathways map of 'core' metabolism (https://www.roche.com/sustainability/philanthropy/science_education/pathways.htm), and the compounds in the KEGG database https://www.genome.jp/kegg/); we note that KEGG contains a number of compounds such as drugs that are not strictly biological molecules. Structures in all three databases were converted into SDF structures with explicit hydrogens using Open Babel (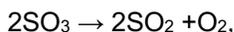http://openbabel.org/), and the numbers of each atom were counted in each of the databases. These are provided in Table S5, together with the fraction that are hydrogen atoms.

## 10. Details of the biologically-based $NH_3$ production model.

Roman numerals refer to elements of the model summarized in Figure 2 of the main text.

(I) $NH_3$ is produced locally in the clouds from atmospheric $N_2$ and $H_2O$ (Table 1, main text) by metabolically active microorganisms (black dots) inhabiting cloud droplets (white circle). $NH_3$ (which is largely confined within the droplet) neutralizes the acid. $NH_3$ production leads to release of $O_2$ which diffuses out of the droplet into the atmosphere, therefore contributing to the anomalous detections of $O_2$ in the clouds.

(II) The production of $NH_3$ raises the droplet pH to ~1 trapping the $SO_2$ and $H_2O$ in droplet as ammonium hydrogen sulfite ($NH_4HSO_3$). The production of sulfite salts in the droplet leads to the formation of a large, semi-solid (and hence non-spherical) Mode 3 particle (white decagon).

(III) The Mode 3 particle settles faster than Mode 2 particles, and falls out of the clouds where ammonium sulfite disproportionates to ammonium sulfate and ammonium sulfide; the latter decomposes to $H_2S$ and $NH_3$, which in turn undergo photochemical reactions to a variety of products.

(IV) Disproportionation and gas release break up the Mode 3 particles into smaller haze particles. Any microorganisms in the Mode 3 particles would have to form spores (black ovals) to survive in this environment. The data available on the stability of ammonium sulfite or ammonium bisulfite in dry conditions is limited, but suggests the ammonium sulfite disproportionates to ammonium sulfate and ammonium sulfide at ~100 °C (32); ammonium sulfide is unstable and dissociates to $NH_3$ and hydrogen sulfide. Survival of spores at 100-120 °C is plausible, based on terrestrial precedent. Some bacteria can even grow at up to 120 °C (in water) (33), and some bacterial spores can survive repeated autoclaving at 135 °C (34); it has even been suggested that some can survive 'ashing' at 420 °C (35).

(V) Some spores can be transported back to the cloud layer by gravity waves, as described by (36). In brief, gravity waves in the atmosphere launched by convective plumes arising in the adjacent (50-55 km and ~18-28 km) convective regions (37–41) can compress atmospheric layers, and allow upwards and downward vertical winds with vertical velocities of ~1 m s$^{-1}$, as measured directly by the Venera landing probes 9 and 10 at the lower haze layer altitudes (and below) with anemometers (42). Metabolically inactive spores that are brought back to the cloud layers would have to act as cloud condensation nuclei (CCN) and survive in conc. $H_2SO_4$ long enough to produce a "first batch" of $NH_3$ to neutralize the acid of the smaller droplet. The mechanism by which this happens is the subject of future work, but possible options are the chemical or photochemical release of $NH_3$ from stored molecules (which chemically neutralizes the condensing $H_2SO_4$), the storage of other poly-bases that could neutralize $H_2SO_4$, possibly forming an ionic liquid phase in which biochemistry could start (e.g. using imidazole derivatives as a base (43)), or that the spores contain highly hygroscopic materials that selectively concentrate water from the environment (perhaps coupled with a selectively sulfuric acid-resistant shell to prevent $H_2SO_4$ from entering the cell as well (36, 44)). See (36) for more details of this transport mechanism.

(VI) The ammonium sulfate particle falls further below the cloud decks, where ammonium sulfate decomposes to $SO_3$, $NH_3$ and $H_2O$, releasing spores into the sub-cloud haze layer as proposed by (36). Ammonium sulfate decomposes at ~200 °C to $NH_3$, water and $SO_3$ (45).

(VII) Spores released at this stage may be unviable (grey ovals), but any surviving could also be transported back to the clouds.

**10.1 Additional commentary on the biologically-based $NH_3$ neutralization of conc. $H_2SO_4$ cloud droplets.**

For completeness we note that $NH_3$ may also be consumed in neutralizing the existing sulfuric acid in the cloud droplets to ammonium hydrogen sulfate ($NH_4HSO_4$). Such scenario could happen if a spore germinates after occupying a pre-formed concentrated sulfuric acid Mode 2 droplet, rather than acting as a CCN and neutralizing acid during the condensation phase (step (V) in the Figure 2 legend). A single cell in an average Mode 2 droplet would take ~6.2 days to convert all the acid in that droplet to ammonium hydrogen sulfate at an $NH_3$ production rate of ~$4 \cdot 10^{-7}$ grams/gram biomass, assuming 80% sulfuric acid. A typical Mode 2 particle would fall down only ~100 m in ~6.2 days (36). However, we note that the exact kinetics of acid neutralization will depend on the complex kinetics of droplet growth and aggregation.

The scenario above assumes that half of the volume of the typical 2 μm Mode 2 particle is occupied by a cell (1.2 μm in diameter) and the second half is filled with a solvent. The other assumptions and results of calculations are as follows:

- Rate of production (from the literature, as mentioned in the main document) = $2.84 \cdot 10^{-8}$ moles $NH_3$/gram biomass
- Diameter of mode 2 particle = 2μm
- Volume ≈ $4.19 \cdot 10^{-12}$ cm$^3$
- Moles of sulfuric acid as pure acid = $5.98 \cdot 10^{-14}$
- Fraction of acid assumed = 80 wt%
- Amount of $NH_3$ required = 0.8 x $5.98 \cdot 10^{-14}$ ≈ $4.8 \cdot 10^{-14}$ moles
- Mass of biomass in the particle assuming 50% occupancy and a density of 1.5 g/cm$^3$ = $3.14 \cdot 10^{-12}$ g
- Seconds for that biomass to make that amount of $NH_3$ = $4.878 \cdot 10^{-14}$ / [$3.14 \cdot 10^{-12}$ x $2.84 \cdot 10^{-8}$] = $5.36 \cdot 10^5$ seconds = 6.2 days.

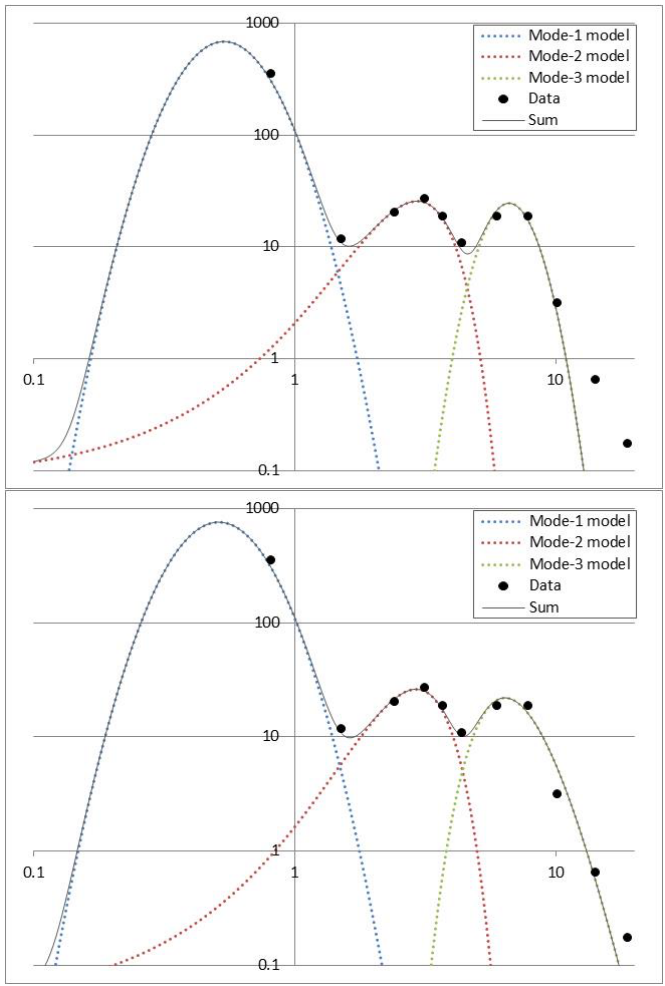

**Fig. S1.** Observed particle abundances as a function of particle size and trimodal model of those abundances. X axis; particle size in microns. Y axis; abundance in particles/cm$^3$/μm. Colored dotted lines are models of the three proposed size modes. Solid line is the total distribution. Black dots are data points from the Pioneer Venus sounder probe. Top Panel; match to trimodal distribution where Mode 1 is log-normal, Mode 2 is normal and Mode 3 is log-normal. Bottom Panel: match to trimodal distribution where Mode 1 is log-normal, Mode 2 is normal and Mode 3 is log-log-normal. We adopted a log-log-normal distribution (i.e. a distribution that is normal for log(log(diameter)), a function that is not defined for diameter < 1) for Mode 3 particles, which fits the observed data better (bottom panel).

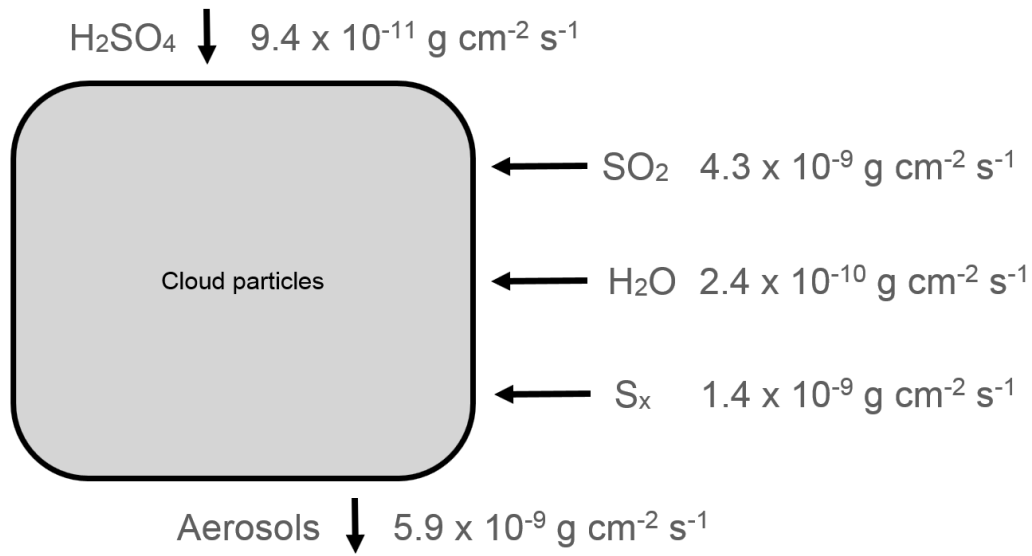

Net Flux: $6.034 \times 10^{-9}$ g cm$^{-2}$ s$^{-1}$ - $5.9 \times 10^{-9}$ g cm$^{-2}$ s$^{-1}$
= $1.34 \times 10^{-10}$ g cm$^{-2}$ s$^{-1}$
N = 51 cm$^{-3}$ vs. 50 cm$^{-3}$, or r = 4.03 μm vs. 4 μm

**Fig. S2.** The overall fluxes of the model.

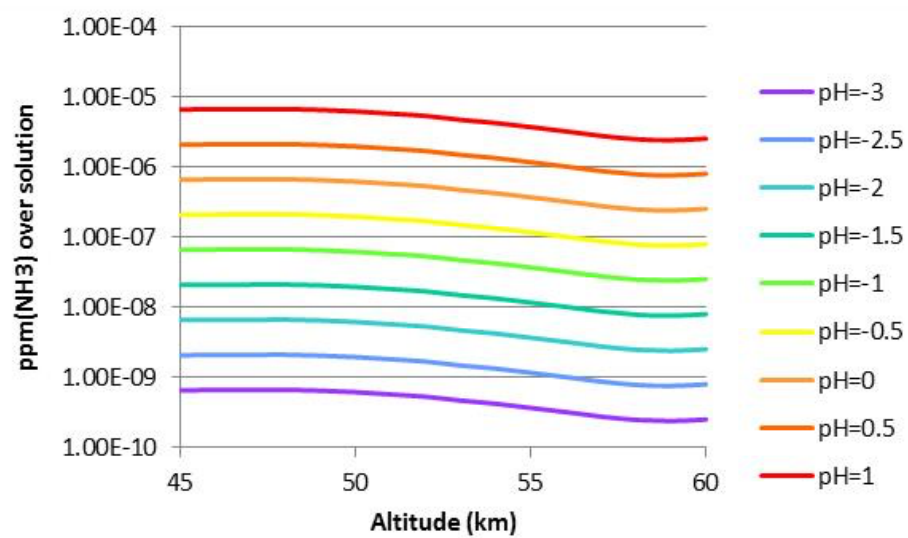

**Fig. S3.** The vapor pressure of sulfuric acid over ammonium sulfate is in the ppt level at 100 °C (46). The detection of vapor phase $H_2SO_4$ at cloud levels on Venus is likely to be due to the gas phase production of $H_2SO_4$ from photochemical oxidation of $SO_2$, as is true on Earth (46).

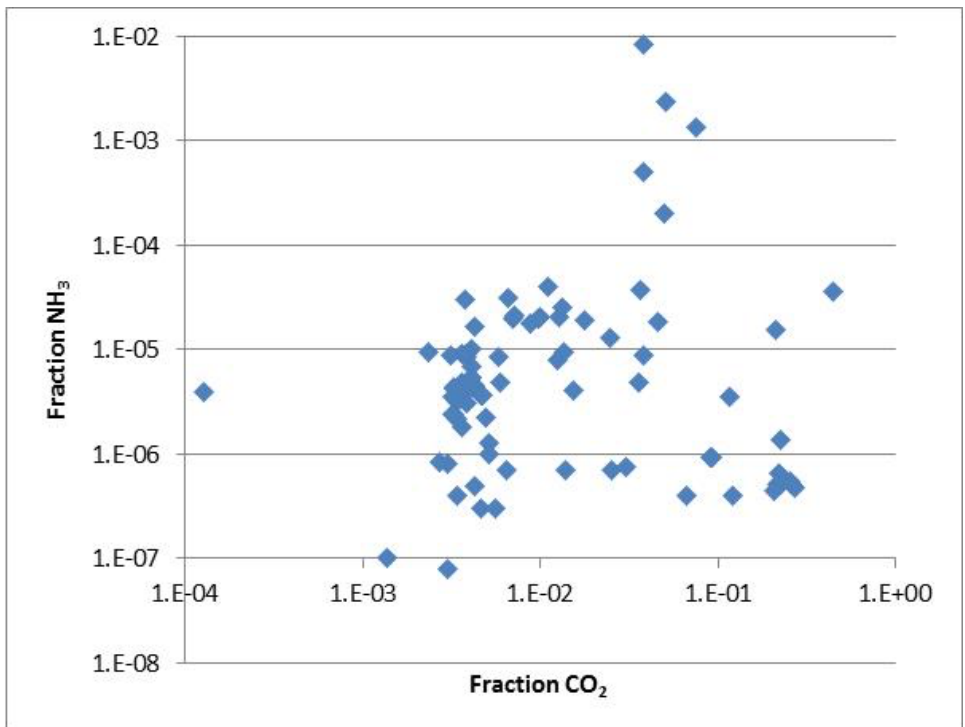

**Fig. S4.** A tabulation of the fraction of gas over 100 °C emitted by terrestrial fumaroles. Fraction of volcanic gas that is $NH_3$ (Y axis) vs fraction that is $CO_2$ (X axis). Data from the following literature sources (47–53).

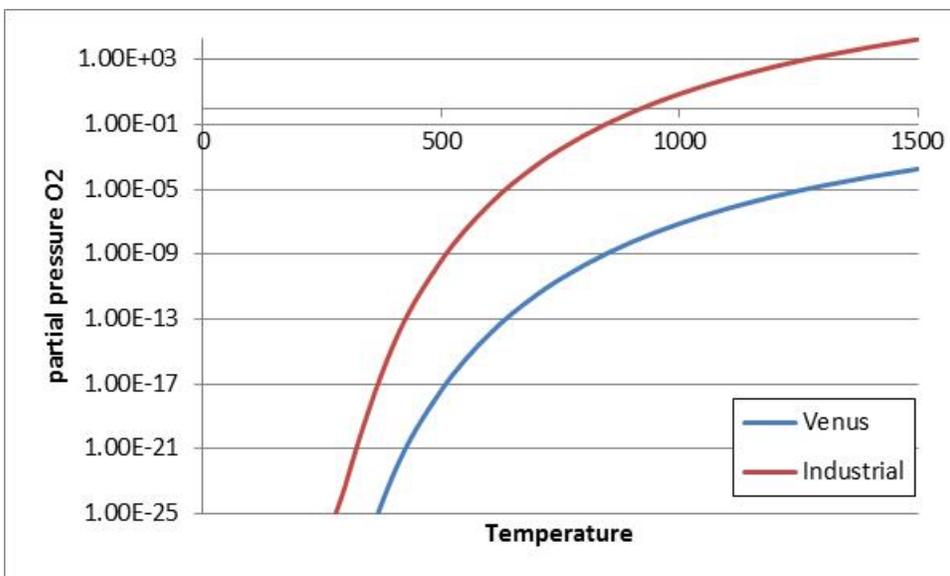

**Fig. S5.** Prediction of the equilibrium abundance of $O_2$ formed from the thermal breakdown of sulfur trioxide, at 1 bar pressure, under Venus and industrial gas abundances. See text for details.

**Table S1.** pKa of SPONCH bases. Only $NH_3$ has a pKa > the pKa of sulfurous acid of 1.71.

| Element | Base equilibrium | pKa |
|---------|------------------|-----|
| N | $NH_3 \leftrightarrow NH_4^+$ | 4.75 |
| N | $N_2H_4 \leftrightarrow N_2H_5^+$ | 5.9 |
| S | $H_2S \leftrightarrow H_3S^+$ | <-11 |
| P | $PH_3 \leftrightarrow PH_4^+$ | -14 |

Hydrazine is more reactive, more unstable to photochemical destruction, and requires more energy to synthesize than $NH_3$, and so we will not consider it further here.

**Table S2.** The distribution of cloud particles at the altitude of 49 km (i.e. lower clouds). Data was taken from the analysis of the Pioneer Venus sounder measurements reported in (2). The particles have a trimodal distribution.

| Mode | Distribution of particle numbers as a function of diameter | Total number per $cm^3$ |
|---|---|---|
| Mode 1 | Log normal | 1200 |
| Mode 2 | Normal | 50 |
| Mode 3 | Log normal | 50 |

**Table S3.** Assessment of the formation of NH₃ in the Venusian environment by lightning.

| Altitude | Temp | Pressure (bar) | Gas phase atoms | | Fraction of N as NH₃ | Bar of NH₃ | moles of NH₃ per cubic meter | Volume of cloud-cloud lightning | Volume cloud-ground lightning | Moles/second |
|---|---|---|---|---|---|---|---|---|---|---|
| | | | H atoms | All other atoms | | | | | | |
| 0 | 735 | 92.1 | 1.266E+24 | 7.347E+27 | 5.115E-12 | 1.649E-11 | 2.984E-10 | 0 | 3.534E+02 | 1.055E-07 |
| 5 | 697 | 66.5 | 9.142E+23 | 5.305E+27 | 5.116E-12 | 1.191E-11 | 2.273E-10 | 0 | 3.534E+02 | 8.032E-08 |
| 10 | 658 | 47.39 | 6.515E+23 | 3.780E+27 | 5.115E-12 | 8.484E-12 | 1.715E-10 | 0 | 3.534E+02 | 6.063E-08 |
| 15 | 621 | 33.04 | 4.684E+23 | 2.636E+27 | 5.610E-12 | 6.488E-12 | 1.390E-10 | 0 | 3.534E+02 | 4.912E-08 |
| 20 | 579 | 22.52 | 3.193E+23 | 1.796E+27 | 5.610E-12 | 4.422E-12 | 1.016E-10 | 0 | 3.534E+02 | 3.591E-08 |
| 25 | 537 | 14.93 | 2.116E+23 | 1.191E+27 | 5.609E-12 | 2.931E-12 | 7.261E-11 | 0 | 3.534E+02 | 2.566E-08 |
| 30 | 495 | 9.851 | 1.338E+23 | 7.858E+26 | 4.935E-12 | 1.701E-12 | 4.573E-11 | 0 | 3.534E+02 | 1.616E-08 |
| 35 | 453 | 5.917 | 8.037E+22 | 4.720E+26 | 4.934E-12 | 1.022E-12 | 3.001E-11 | 0 | 3.534E+02 | 1.060E-08 |
| 40 | 416 | 3.501 | 4.755E+22 | 2.793E+26 | 4.932E-12 | 6.044E-13 | 1.933E-11 | 0 | 3.534E+02 | 6.831E-09 |
| 45 | 383 | 1.979 | 2.687E+22 | 1.579E+26 | 4.931E-12 | 3.416E-13 | 1.186E-11 | 7.854E+01 | 0 | 9.318E-08 |
| 50 | 348 | 1.066 | 1.807E+21 | 8.501E+25 | 9.608E-15 | 3.585E-16 | 1.370E-14 | 7.854E+01 | 0 | 1.076E-10 |
| 55 | 300 | 0.5314 | 2.930E+20 | 4.237E+25 | 3.306E-16 | 6.149E-18 | 2.727E-16 | 7.854E+01 | 0 | 2.142E-12 |
| 60 | 263 | 0.2357 | 6.444E+19 | 1.879E+25 | 4.031E-17 | 3.325E-19 | 1.682E-17 | 7.854E+01 | 0 | 1.321E-13 |
| 65 | 243 | 0.0976 | 2.669E+19 | 7.787E+24 | 4.027E-17 | 1.376E-19 | 7.534E-18 | 7.854E+01 | 0 | 5.917E-14 |
| 70 | 230 | 0.0369 | 1.009E+19 | 2.942E+24 | 4.026E-17 | 5.200E-20 | 3.008E-18 | 7.854E+01 | 0 | 2.362E-14 |

**Table S4.** Calculation of the thermodynamic equilibrium concentration of $NH_3$ and $O_2$ if the entire atmosphere was 'shocked' to high temperature and quenched at 1200 K. All properties were evaluated using the Chemkin® Collection and STANJAN (https://navier.engr.colostate.edu/code/code-4/index.html, 2019 David Dandy).

### Chemical Equilibrium Results

|  | Initial State | Equilibrium State |
|---|---|---|
| Pressure (atm) | 9.8692E+01 | 9.8692E+01 |
| Temperature (K) | 1.2000E+03 | 1.2000E+03 |
| Volume (cm³/g) | 2.2920E+01 | 2.2919E+01 |
| Enthalpy (erg/g) | -7.7578E+10 | -7.7578E+10 |
| Internal Energy (erg/g) | -7.9870E+10 | -7.9870E+10 |
| Entropy (erg/g K) | 5.5354E+07 | 5.5354E+07 |

|  | Initial State | | Equilibrium State | |
|---|---|---|---|---|
|  | mole fraction | mass fraction | mole fraction | mass fraction |
| CO2 | 9.6973E-01 | 9.8040E-01 | 9.6973E-01 | 9.8040E-01 |
| N2 | 2.9992E-02 | 1.9301E-02 | 2.9992E-02 | 1.9301E-02 |
| H2O | 2.9992E-05 | 1.2412E-05 | 3.0489E-05 | 1.2618E-05 |
| SO2 | 1.4996E-04 | 2.2069E-04 | 1.4996E-04 | 2.2069E-04 |
| CO | 9.9972E-05 | 6.4328E-05 | 1.0047E-04 | 6.4649E-05 |
| H2 | 4.9986E-07 | 2.3149E-08 | 2.3128E-09 | 1.0711E-10 |
| COS | 0.0000E+00 | 0.0000E+00 | 4.5671E-11 | 6.3028E-11 |
| SO | 0.0000E+00 | 0.0000E+00 | 7.3588E-10 | 8.1251E-10 |
| S2 | 0.0000E+00 | 0.0000E+00 | 3.9969E-15 | 5.8881E-15 |
| H2S | 0.0000E+00 | 0.0000E+00 | 3.2978E-14 | 2.5818E-14 |
| O2 | 0.0000E+00 | 0.0000E+00 | 2.7812E-10 | 2.0444E-10 |
| CH4 | 0.0000E+00 | 0.0000E+00 | 1.6541E-25 | 6.0962E-26 |
| NH3 | 0.0000E+00 | 0.0000E+00 | 3.6998E-16 | 1.4475E-16 |
| NO | 0.0000E+00 | 0.0000E+00 | 1.5322E-09 | 1.0561E-09 |
| NO2 | 0.0000E+00 | 0.0000E+00 | 9.0913E-15 | 9.6082E-15 |

**Table S5.** Counts of atoms in all of three databases of biological molecules, the natural products database compiled by (31), the Roche metabolic map database of 'core' metabolism, and KEGG database.

| Dataset | Number of molecules | Percentage of atoms that are H | Average molecular weight | C | H | N | O | P | S | Other Elements |
|---|---|---|---|---|---|---|---|---|---|---|
| Roche map metabolites | 699 | 49.34% | 312.97 | 9056 | 14337 | 1205 | 4028 | 335 | 78 | 16 |
| KEGG | 15543 | 49.01% | 392.12 | 296944 | 403473 | 22376 | 89934 | 3221 | 2707 | 4643 |
| Natural Products Database | 204053 | 49.16% | 481.43 | 5290405 | 6769512 | 137557 | 1545192 | 1292 | 11534 | 13551 |

**Table S6.** Boundary conditions for the photochemical model of the Venusian atmosphere.

| $CO_2$ | $N_2$ | $SO_2$ | $H_2O$ | CO | OCS | HCl | $H_2S$ | NO | $H_2$ |
|---|---|---|---|---|---|---|---|---|---|
| 96% | 3% | 150 ppm | 30 ppm | 20 ppm | 5 ppm | 500 ppb | 10 ppb | 5.5 ppb | 3 ppb |